\documentstyle[sprocl,epsf]{article}

 \catcode`@=11
\def\@cite#1#2{\unskip\nobreak\relax
    \def\@tempa{$\m@th^{\hbox{\the\scriptfont0 #1}}$}%
    \futurelet\@tempc\@citexx}
\def\@citexx{\ifx.\@tempc\let\@tempd=\@citepunct\else
    \ifx,\@tempc\let\@tempd=\@citepunct\else
    \ifx;\@tempc\let\@tempd=\@citepunct\else
    \ifx:\@tempc\let\@tempd=\@citepunct\else
\let\@tempd=\@tempa\fi\fi\fi\fi\@tempd}
\def\@citepunct{\@tempc\edef\@sf{\spacefactor=\the\spacefactor\relax}\@tempa
    \@sf\@gobble}
 
\def\citenum#1{{\def\@cite##1##2{##1}\cite{#1}}}
\def\citea#1{\@cite{#1}{}}
 
\newcount\@tempcntc
\def\@citex[#1]#2{\if@filesw\immediate\write\@auxout{\string\citation{#2}}\fi
  \@tempcnta\z@\@tempcntb\m@ne\def\@citea{}\@cite{\@for\@citeb:=#2\do
    {\@ifundefined
       {b@\@citeb}{\@citeo\@tempcntb\m@ne\@citea\def\@citea{,}{\bf ?}\@warning
       {Citation `\@citeb' on page \thepage \space undefined}}%
    {\setbox\z@\hbox{\global\@tempcntc0\csname b@\@citeb\endcsname\relax}%
     \ifnum\@tempcntc=\z@ \@citeo\@tempcntb\m@ne
       \@citea\def\@citea{,}\hbox{\csname b@\@citeb\endcsname}%
     \else
      \advance\@tempcntb\@ne
      \ifnum\@tempcntb=\@tempcntc
      \else\advance\@tempcntb\m@ne\@citeo
      \@tempcnta\@tempcntc\@tempcntb\@tempcntc\fi\fi}}\@citeo}{#1}}
\def\@citeo{\ifnum\@tempcnta>\@tempcntb\else\@citea\def\@citea{,}%
  \ifnum\@tempcnta=\@tempcntb\the\@tempcnta\else
   {\advance\@tempcnta\@ne\ifnum\@tempcnta=\@tempcntb \else \def\@citea{--}\fi
    \advance\@tempcnta\m@ne\the\@tempcnta\@citea\the\@tempcntb}\fi\fi}
\catcode`@=12

\catcode`@=11
\def\subsection{\@startsection{subsection}{2}{\z@}{-3.25ex plus -1ex minus 
    -.2ex}{1.5ex plus .2ex}{\rm }}
\def\subsubsection{\@startsection{subsubsection}{3}{\z@}{-3.25ex plus -1ex minus -.2ex}{1.5ex plus .2ex}{\it }}
\catcode`@=12

\def\lsim{\mathrel{\raise.2ex\hbox{$<$}\hskip-.8em\lower.9ex\hbox{$\sim$}}}
\def\gsim{\mathrel{\raise.2ex\hbox{$>$}\hskip-.8em\lower.9ex\hbox{$\sim$}}}
\let\alt=\lsim
\let\agt=\gsim

\begin{document}
\thispagestyle{empty}

\renewcommand{\thefootnote}{\fnsymbol{footnote}}

\vglue-.7in

\font\fortssbx=cmssbx10 scaled \magstep1
\hbox to \hsize{
\includegraphics{/NextLibrary/TeX/tex/inputs/uwlogo.ps}
\hskip.25in \raise.05in\hbox{\fortssbx University of Wisconsin - Madison}
\hfill\vbox{\hbox{\bf MADPH-98-1088}
            \hbox{October 1998}} }

\bigskip

\title{\uppercase{Lectures on Neutrino Astronomy:\\ Theory and Experiment}\footnote{Lectures presented at the TASI School, July 1998.}}

\author{\unskip\medskip\uppercase{Francis Halzen}}

\address{Department of Physics, University of Wisconsin, Madison, WI 53706}

\maketitle\abstracts{
1. Overview of neutrino astronomy: multidisciplinary science.\\
2. Cosmic accelerators: the highest energy cosmic rays.\\
3. Neutrino beam dumps: supermassive black holes and gamma ray bursts.\\
4. Neutrino telescopes: water and ice.\\
5. Indirect dark matter detection.\\
6. Towards kilometer-scale detectors.
}

\section{Neutrino Astronomy: Multidisciplinary Science}

Using optical sensors buried in the deep clear ice or deployed in deep ocean and lake waters,  neutrino astronomers are attacking major problems in astronomy, astrophysics, cosmic ray physics and particle physics by commissioning first generation neutrino telescopes.  Planning is already underway to instrument a cubic volume of ice or water, 1 kilometer on the side, as a detector of neutrinos in order to reach the effective telescope area of 1 kilometer squared which is, according to estimates covering a wide range of scientific objectives, required to address the most fundamental questions. This infrastructure provides unique opportunities for yet more interdisciplinary science covering the geosciences and biology.

Among the many problems which high energy neutrino telescopes will address are the origin of cosmic rays, the engines which power active galaxies, the nature of gamma ray bursts, the search for the annihilation products of halo cold dark matter (WIMPS, supersymmetric particles(?)), galactic supernovae and, possibly, even the structure of Earth's interior. Coincident experiments with Earth- and space-based gamma ray observatories, cosmic ray telescopes and gravitational wave detectors such as LIGO can be contemplated. With high-energy neutrino astrophysics we are poised to open a new window into space and back in time to the highest-energy processes in the Universe.

``And the estimate of the primary neutrino flux may be too low, since regions that produce neutrinos abundantly may not reveal themselves in the types of radiation yet detected" Greisen states in his 1960 review\cite{greisen}. He also establishes that the natural scale of a deep underground neutrino detector is 15\,m! The dream of neutrino astronomers is the same today, but experimental techniques are developed with the ultimate goal to deploy kilometer-size instruments.

I will first introduce high energy neutrino detectors as astronomical telescopes using Fig.\,1. The figure shows the diffuse flux of photons as a function of their energy and wavelength, from radio-waves to the high energy gamma rays detected with satellite-borne detectors\cite{turner}. The data span 19 decades in energy. Major discoveries have been historically associated with the introduction of techniques for exploring new wavelenghts. All of the discoveries were surprises; see Table\,1. The primary motivation for commissioning neutrino telescopes is to cover the uncharted territory in Fig.\,1: wavelengths smaller than $10^{-14}$\,cm, or energies in excess of 10\,GeV. This exploration has already been launched by truly pioneering observations using air Cerenkov telescopes\cite{weekes}. Larger space-based detectors as well as cosmic ray facilities with sensitivity above $10^7$\,TeV, an energy where charged protons point back at their sources with minimal deflection by the galactic magnetic field, will be pursuing similar goals. Could the high energy skies in Fig.\,1 be empty? No, cosmic rays with energies exceeding $10^8$\,TeV have been recorded\cite{cronin}.

\begin{figure}[h]
\centering
\hspace{0in}\epsfxsize=3.7in\epsffile{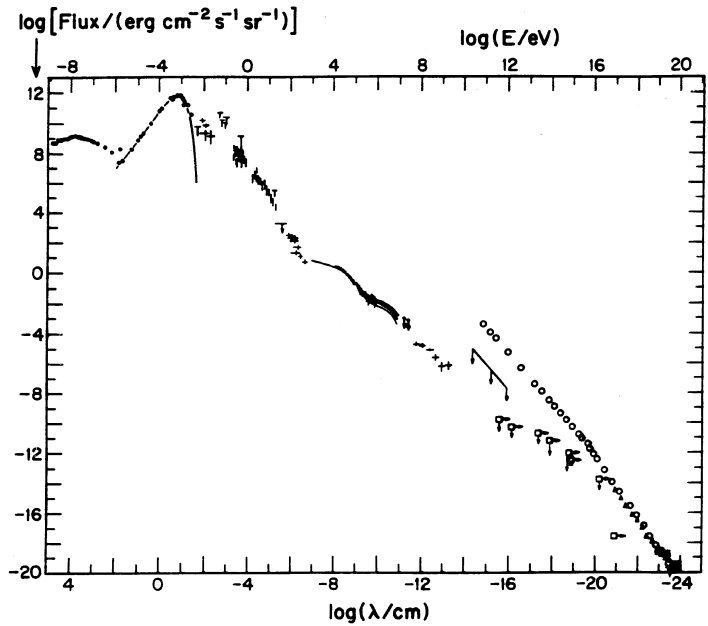}

\caption{Flux of gamma rays as a function of wavelength and photon energy. In the TeV--EeV energy range the anticipated fluxes are dwarfed by the cosmic ray flux which is also shown in the figure.}
\end{figure}

\begin{table}[t]
\caption{New windows on the Universe}
\bigskip
\tabcolsep=.2em
\centering
\begin{tabular}{|l@{\hspace{-.5em}}c@{\hspace{-.5em}}c|}
\hline
\quad\bf Telescope& \bf Intended use& \bf Actual results\\
\hline
optical (Galileo)& navigation& moons of Jupiter\\ 
radio (Jansky)& noise& radio galaxies\\
optical (Hubble)& nebulae& expanding Universe\\
microwave (Penzias-Wilson)& noise& 3K cosmic background\\
X-ray (Giacconi\dots)& moon& neutron star acc.\ binaries\\
radio (Hewish, Bell)& scintillations& pulsars\\
$\gamma$-ray (???)& thermonuclear explosions& $\gamma$-ray bursts\\
\hline
\end{tabular}
\end{table}

Exploring this wide energy region with neutrinos does have the definite advantage that they can, unlike high energy photons and nuclei, reach us, essentially without attenuation in flux, from the largest red-shifts. Gamma rays come from a variety of objects, both galactic (supernova remnants such as the Crab Nebula) and extragalactic (active galaxies), with energies up to at least 30 TeV, but they are absorbed by extragalactic infrared radiation in distances less than 100 Mpc. Photons of TeV energy and above are efficiently absorbed by pair production of electrons on background light above a threshold
\begin{equation}
4E\epsilon > (2m_e)^2 \,,
\end{equation}
where $E$ and  $\epsilon$ are the energy of the accelerated and background photon in the c.m.\ system, respectively. Therefore TeV photons are absorbed on infrared light, PeV photons ($10^3$\,TeV) on the cosmic microwave background and EeV ($10^{18}$\,eV) on radiowaves. It is likely that absorption effects explain why Markarian 501, at a distance of barely over 100\,Mpc the closest blazar on the EGRET list of sources, produces the most prominent TeV signal. Although one of the closest active galaxies, it is one of the weakest; the reason that it is detected whereas other, more distant, but more powerful, AGN are not, must be that the TeV gamma rays suffer absorption in intergalactic space through the interaction with background infra-red light\cite{salomon}. Absorption most likely provides the explanation why much more powerful quasars such as 3C279 at a redshift of 0.54 have not been identified as TeV sources.

Cosmic rays are accelerated to energies as high as $10^8$\,TeV. Their range in intergalactic space is also limited by absorption, not by infrared light but by the cosmic microwave radiation. Protons interact with background light by the production of the $\Delta$-resonance just above the threshold for producing pions:
\begin{equation}
4E_p\epsilon > \left(m_\Delta^2 - m_p^2\right) \,.
\label{eq:threshold}
\end{equation}
The dominant energy loss of protons of $\sim$500\,EeV energy and above, is photoproduction of the $\Delta$-resonance on cosmic microwave photons. The Universe is therefore opaque to the highest energy cosmic rays, with an absorption length of only tens of megaparsecs when their energy exceeds $10^8$\,TeV. Lower energy protons, below threshold (\ref{eq:threshold}), do not suffer this fate. They cannot be used for astronomy however because their direction is randomized in the microgauss magnetic field of our galaxy. Of all high-energy particles, only neutrinos convey directional information from the edge of the Universe and from the hearts of the most cataclysmic high energy processes.

Although Nature is clearly more imaginative than scientists, as illustrated in Table\,1, active galactic nuclei (AGN) and gamma ray bursts (GRB) must be considered well-motivated sources of neutrinos simply because they are the sources of the most energetic photons. They may also be the accelerators of the highest energy cosmic rays. If they are, their neutrino flux can be calculated in a relatively model-independent way because the proton beams will photoproduce pions and, therefore, neutrinos on the high density of photons in the source. We have a beam dump configuration where both the beam and target are constrained by observation: by cosmic ray observations for the proton beam and by astronomical observations for the photon target. AGN and GRB have served as the most important ``gedanken experiments" by which we set the scale of future neutrino telescopes. We will show that order 100 detected neutrinos are predicted per year in a high energy neutrino telescope with an effective area of 1\,km$^2$. Their energies cluster in the vicinity of 100\,TeV for GRB and several 100\,PeV for neutrinos originating in AGN jets. For the latter, even larger fluxes of lower energy energy neutrinos may emanate from their accretion disks.

Neutrino telescopes can do particle physics. This is often illustrated by their capability to detect the annihilation into high energy neutrinos of neutralinos, the lightest supersymmetric particle which may constitute the dark matter. This will be discussed in some detail in the context of the AMANDA detector in Section~5. Also, with cosmological sources such as active galaxies and GRB we will be observing $\nu_e$ and $\nu_\mu$ neutrinos over a baseline of $10^3$\,Megaparsecs. Above 1\,PeV these are absorbed by charged-current interactions in the Earth before reaching the opposite surface.  In contrast, the Earth never becomes opaque to $\nu_\tau$ since the $\tau^-$ produced in a charged-current $\nu_\tau$ interaction decays back into $\nu_\tau$ before losing significant energy.  This penetration of tau neutrinos through the Earth above $10^2$\,TeV provides an experimental signature for neutrino oscillations. The appearance of a $\nu_{\tau}$ component in a pure $\nu_{e,\mu}$ would be evident as a flat angular dependence of a source intensity at the highest neutrino energies.  Such a flat zenith angle dependence for the farthest sources would indicate tau neutrino mixing with a sensitivity to $\Delta m^2$ as low as $10^{-17}$\,eV$^2$. With neutrino telescopes we will also search for ultrahigh-energy neutrino signatures from topological defects and magnetic monopoles; for properties of neutrinos such as mass, magnetic moment, and flavor-oscillations; and for clues to entirely new physical phenomena. The potential of neutrino ``telescopes" to do particle physics is evident.

We start with a discussion of how Nature may accelerate subnuclear particles to macroscopic energies of more than 10\,Joules. We discuss cosmic beam dumps next. We conclude with a status report on the deployment of the first-generation neutrino telescopes. 

\section{Cosmic Accelerators}

\subsection{The Machines}

Cosmic rays form an integral part of our galaxy. Their energy density is qualitatively similar to that of photons, electrons and magnetic fields. It is believed that most were born in supernova blast waves. Their energy spectrum can be understood, up to perhaps 1000\,TeV, in terms of acceleration by supernova shocks exploding into the interstellar medium of our galaxy. Although the slope of the cosmic ray spectrum abruptly increases at this energy, particles with energies in excess of $10^{8}$\,TeV have been observed and cannot be accounted for by this mechanism. The break in the spectrum, usually referred to as the ``knee\rlap,'' can be best exhibited by plotting the flux multiplied by an energy dependent power $E^{2.75}$; see Fig.\,2. 

\begin{figure}[t]
\centering\leavevmode
\epsfxsize=3.5in\epsffile{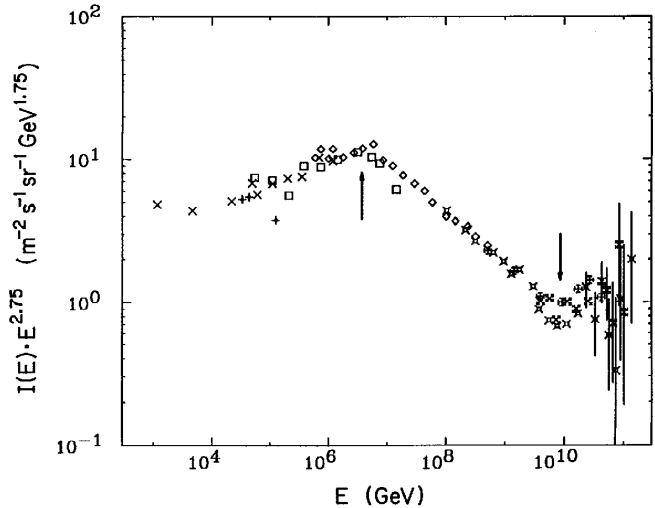}

\caption{Flux of high energy cosmic rays after multiplication by a factor $E^{2.75}$. Arrows point at structure in the spectrum near 1~PeV, the ``knee\rlap," and 10~EeV, the ``ankle\rlap."}
\end{figure}

The failure of supernovae to accelerate cosmic rays above 1000\,TeV energy can be essentially understood on the basis of dimensional analysis. It is sensible to assume that, in order to accelerate a proton to energy $E(=pc)$, the size $R$ of the accelerator must be larger than the gyroradius of the particle in the accelerating field $B$:
\begin{equation}
R > R_{\rm gyro} = {E\over B} \,,
\end{equation}
in units $e=c=1$. This yields a maximum energy
\begin{equation}
E < BR, 
\end{equation}
or
\begin{equation}
\left[ E_{\rm max}\over 10^5\,\rm TeV \right] = 
\left[ B \over 3\times 10^{-6}\rm G \right] \left[ R\over 50\rm\,pc \right]\;.
\end{equation}
Therefore particles moving at the speed of light $c$ reach energies up to a maximum value $E_{\rm max}$ which must be less than $10^5$\,TeV for the values of $B$ and $R$ characteristic for a supernova shock. Realistic modelling introduces inefficiencies in the acceleration process and yields a maximum energy which is typically two orders of magnitude smaller than the value obtained by dimensional analysis; see later. One therefore identifies the ``knee" with the sharp cutoff associated with particles accelerated by supernovae.

Cosmic rays with energy in excess of $10^8$\,TeV have been observed, some five orders of magnitude in energy above the supernova cutoff. Where and how they are accelerated undoubtedly represents one of the most challenging problems in cosmic ray astrophysics and one of the oldest unresolved puzzles in astronomy. In order to beat dimensional analysis, one must accelerate particles over larger distances $R$, or identify higher magnetic fields $B$.

Although imaginative arguments actually do exist to avoid this conclusion, it is generally believed that our galaxy is too small and its magnetic field too weak to accelerate the highest energy cosmic rays. Furthermore, those with energy in excess of $10^7$\,TeV have gyroradii larger than our galaxy and should point back at their sources. Their arrival directions fail to show any correlation to the galactic plane, suggesting extra-galactic origin. Searching the sky beyond our galaxy, the central engines of active galactic nuclei and the enigmatic gamma ray bursts stand out as the most likely sites from which particles can be hurled at Earth with joules of energy. The idea is rather compelling because bright AGN and GRB are also the sources of the highest energy photons detected with satellite and air Cherenkov telescopes.

\begin{table}[h]
\caption{}
\bigskip
\centering

\hspace{0in}\epsfxsize=3in\epsffile{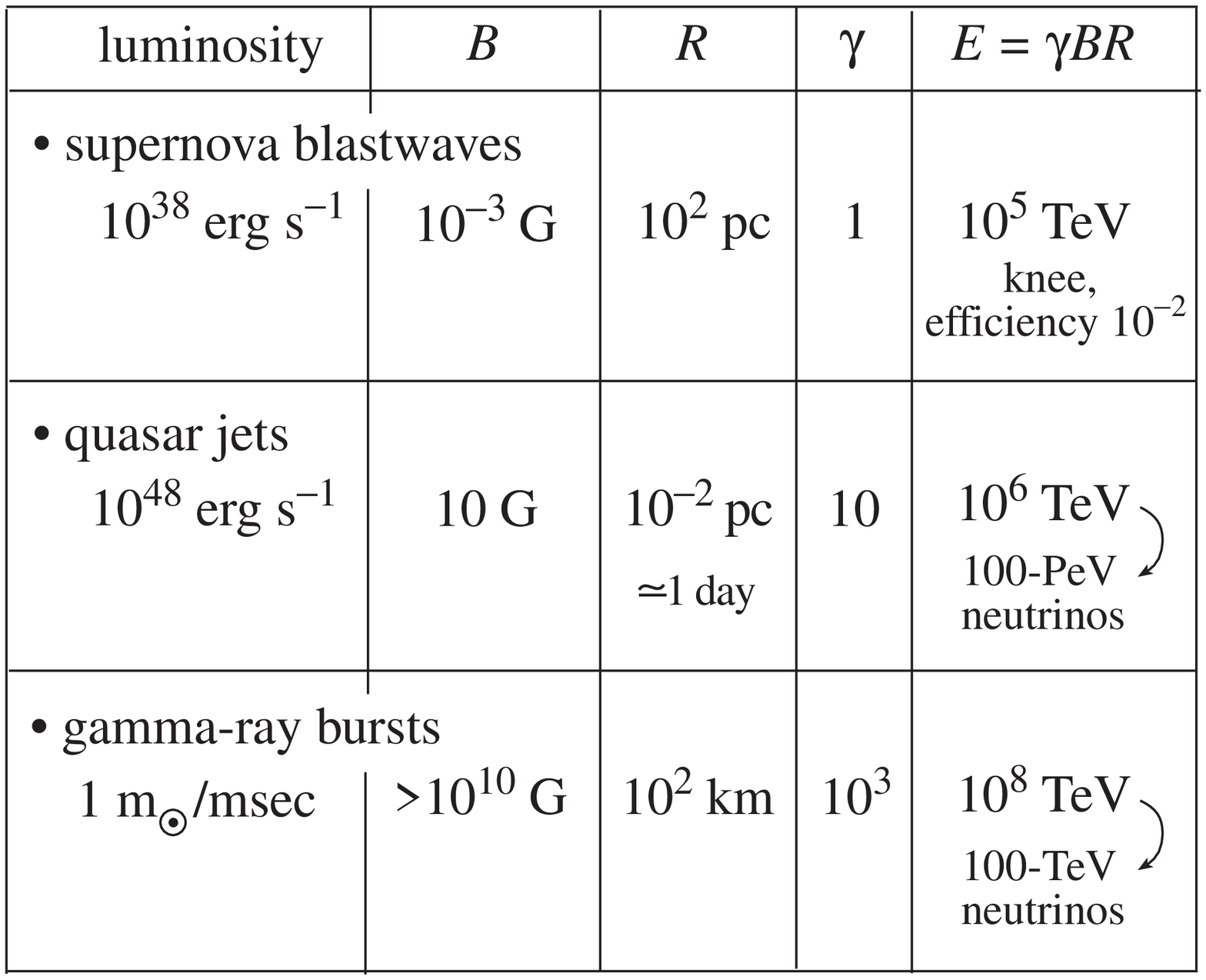}
\end{table}

The jets in AGN consists of beams of electrons and protons accelerated by tapping the rotational energy of the black hole. The black hole is the source of the phenomenal AGN luminosity, emitted in a multi-wavelength spectrum extending from radio up to gamma rays with energies in excess of 20\,TeV. In a jet, $BR$-values in excess of $10^6$\,TeV can be reached with fields of tens of Gauss extending over sheets of shocked material in the jet of dimension $10^{-2}$\,parsecs; see Table\,2. The size of the accelerating region is deduced from the duration of the high energy emission which occurs in bursts lasting days, sometimes minutes; see Figs.\,3, 4. Near the supermassive black hole Nature does not only construct a beam, the beam is a beam of smaller accelerating regions. AGN create accelerators more powerful than Fermilab about once a day! The $\gamma$-factor in Table\,2 reminds us that in the most spectacular sources the jet is beamed in our direction, thus increasing the energy and reducing the duration of the emission in the observer's frame. As previously mentioned, accelerated protons interacting with the ambient light are assumed to be the source of secondary pions which produce the observed gamma rays and, inevitably, neutrinos\cite{agn}.

\begin{figure}[t]
\centering
\epsfxsize=2.5in\hspace{0in}\epsffile{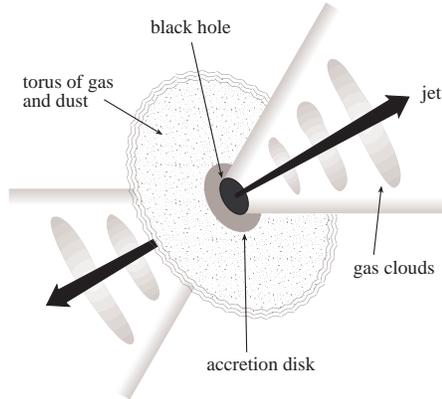}

\caption{Active galaxy with accretion disk and a pair of jets. The galaxy is powered by a central super-massive black hole ($\sim 10^9 M_\odot$). Particles, accelerated in shocks in the disk or the jets, interact with the high density of ambient photons ($\sim 10^{14}$/cm$^3$).}
\end{figure}

In this context, GRB and AGN jets are similar objects. GRB are somehow associated with neutron stars or solar mass black holes. Characteristic fields in excess of $10^{10}$\,Gauss are concentrated in a fireball of size $10^2$\,kilometers, which is opaque to light. The relativistic shock ($\gamma \simeq 10^3$) which dilutes the fireball to the point where the gamma ray display occurs, will also accelerate protons. These interact with the observed light to produce neutrinos\cite{grb}; see Table\,2.

\subsection{The Blueprint: Shock Acceleration}

Cosmic and Earth-based accelerators operate by different mechanisms. In space electric fields of freely moving particles short, and magnetic fields are generally disorganized. Particles gain energy in collisions. A particle of mass $m$ and velocity $v$ colliding head-on with a stellar cloud of mass $M$ and velocity $u$, gains a kinetic energy:
\begin{equation}
{\Delta E\over E}(u,v) \sim {E_{\rm after} - E_{\rm before}\over{1\over2} mv^2}
\sim {u\over v} \left(1 + {u\over v}\right) \,.
\label{gain}
\end{equation}
In order to find the net gain in energy we have to average over all possible directions of the particle-cloud collision. In a one-dimensional (and, note, non-relativistic) world the particle encounters $(v+u)/v$ clouds, head-on, and collides with a smaller number $(v-u)/v$ coming from the opposite direction. The argument is familiar from the Doppler shift. The net gain in energy is given by:
\begin{equation}
\left(\Delta E\over E\right)_{\rm net} \sim {\Delta E\over E}(u,v) \left(u+v\over v\right) + {\Delta E\over E}(-u,v) \left(v-u\over v\right)
\sim  \left(u\over v\right)^2 \,.
\label{netgain}
\end{equation}
The gain is very small because only the quadratic term $\left(u/v\right)^2$ survives. In astrophysical situations $u$ will be typically much smaller than the relativistic particle velocity $v$. It is clear that averaging over ``good" and ``bad" collisions is the origin of the cancellation of the linear contribution $u/v$ to the gain in energy.

The key is to find an environment where particles only undergo ``good" collisions: a shock, for instance the shock expanding into the interstellar medium produced by a supernova explosion. Acceleration in shocks is referred to as first-order Fermi acceleration, for obvious reasons. In astrophysical shocks the collisions are really magnetic interactions: with magnetic irregularities upstream and turbulence downstream. The particles riding the shockwave collide head-on with both! The highest energy particles will have crossed the shock many times. The details are not simple\cite{longair}. Fortunately Nature has provided us with many examples of shocks in action. For example, solar particles produced with MeV energy in nuclear collisions are observed with GeV energy as a result of shock acceleration at the surface.

Unlike the typical mono-energetic beam produced by a synchrotron, shocks produce power law spectra of high energy particles,
\begin{equation}
dN/dE\propto E^{-(\gamma+1)} \,. \label{eq:dN/dE}
\end{equation}
The observed high energy cosmic ray spectrum at Earth is characterized by $\gamma\simeq 1.7$.  A cosmic accelerator in which the dominant mechanism is first order
diffusive shock acceleration, will produce a spectrum with $\gamma \sim 1 +\epsilon$, where $\epsilon$ is a small number.  The observed cosmic ray spectrum is steeper than the accelerated spectrum because of the energy dependence of their diffusion in the galaxy: high energy ones more readily escape confinement from the magnetic bottle formed by the galaxy. In highly relativistic shocks $\epsilon$ can take a negative value.

First-order Fermi acceleration at supernova blast shocks offers a very attractive model for a galactic cosmic accelerator, providing the right power and spectral shape. Acceleration takes time, however, because the energy gain occurs gradually as a particle near the shock scatters back and forth across the front, gaining energy with each transit. The finite lifetime of the shock thus limits the maximum energy a particle can achieve at a particular supernova shock.  The acceleration rate is\cite{pr}
\begin{equation}
{\Delta E\over\Delta t} = K {u^2\over c} \, ZeB < ZeBc \,,
\label{eq:acc-rate}
\end{equation}
where $u$ is the shock velocity, $Ze$ the charge of the particle being accelerated and $B$
the ambient magnetic field.  The numerical constant $K\sim 0.1$ is an efficiency factor which depends on the details of diffusion in the vicinity of the shock such as the efficiency by which power in the shock is converted into the actual acceleration of particles. The maximum energy reached is
\begin{equation}
E = {K\over c} (ZeB\,R u) < ZeB\,R \,.
\label{eq:Emax}
\end{equation}
The crucial time scale used to convert Eq.\,(\ref{eq:acc-rate}) into this limiting energy is $\Delta t\sim R/u$, where $\Delta t \sim 1000$\,yrs for the free expansion phase of a supernova and $R$ is the dimension of the blastwave. This result agrees with Eq.\,(4). Using Eq.\,(\ref{eq:Emax}) we ascertain that $E_{\rm max}$ can only reach energies of $\alt 10^{3}$\,TeV${}\times Z$ for a galactic field $B \sim 3\mu$Gauss, $K\sim 0.1$ and $u/c \sim 0.1$. Even ignoring all pre-factors the energy can never exceed $10^5$\,TeV by dimensional analysis. Cosmic rays with energy in excess of $10^8$\,TeV have been observed and the acceleration mechanism leaves a large gap of some five orders of magnitude that cannot be explained by the ``standard model'' of cosmic ray origin. To reach a higher energy one has to dramatically increase $B$ and/or $R$. This argument is difficult to beat --- it is basically dimensional. Even the details do not matter; elementary electromagnetism is sufficient to identify the EMF of the accelerator or even the Lorentz force in the form of Eq.\,(\ref{eq:Emax}).

First-order Fermi acceleration is also believed to be the origin of the very high energy particles produced near the supermassive black holes in active galaxies and in the explosive release of a solar mass in gamma ray bursts.

\section {Neutrinos from Cosmic Beam Dumps}

Cosmic neutrinos, just like accelerator neutrinos, are made in beam dumps. A beam of accelerated protons is dumped into a target where they produce pions in collisions with nuclei. Neutral pions decay into gamma rays and charged pions into muons and neutrinos. All this is standard particle physics and, in the end, roughly equal numbers of secondary gamma rays and neutrinos emerge from the dump. In man-made beam dumps the photons are absorbed in the dense target; this may not be the case in an astrophysical system where the target material can be more tenuous. Also, the target may be photons rather than nuclei. For instance, with an ambient photon density a million times larger than the sun, approximately $10^{14}$ per cm$^3$, particles accelerated in AGN jets may meet more photons than nuclei when losing energy. Examples of cosmic beam dumps are tabulated in Table\,3. They fall into two categories. Neutrinos produced by the cosmic ray beam are, of course, guaranteed and calculable\cite{pr}. We know the properties of the beam and the various targets: the atmosphere, the hydrogen in the galactic plane and the CMBR background. Neutrinos from AGN and GRB  are not guaranteed, though both represent good candidate sites for the acceleration of the highest energy cosmic rays. That they are also the sources of the highest energy photons reinforces this association.

\begin{table}[h]
\caption{Cosmic Beam Dumps}
\bigskip
\centering
\tabcolsep=1em
\begin{tabular}{|c|c|}
\hline
\bf Beam& \bf Target\\
\hline
cosmic rays& atmosphere\\
cosmic rays& galactic disk\\
cosmic rays& CMBR\\
AGN jets& ambient light, UV\\
shocked protons& GRB photons\\
\hline
\end{tabular}
\end{table}

Our main point will be that the rate of neutrinos produced by $p\gamma$ interactions in gamma ray bursts and active galactic nuclei is essentially dictated by the observed energetics of the source. In astrophysical beam dumps, like AGN and GRB, typically one neutrino and one photon is produced per accelerated proton\cite{pr}. The accelerated protons and photons are, however, more likely to suffer attenuation in the source before they can escape. So, a hierarchy of particle fluxes emerges with protons\,$<$\,photons\,$<$\,neutrinos. Using these associations, one can constrain the energy and luminosity of the accelerator from the gamma-ray and cosmic ray observations, and subsequently anticipate the neutrino fluxes. These calculations represent the basis for the construction of kilometer-scale detectors as the goal of neutrino astronomy.   

Below follow 3 estimates of the luminosity of the Universe in high energy radiation. 

\begin{enumerate}

\item From the observed injection rate $\dot E = 4\times10^{44}\rm\, erg \, Mpc^{-3}\, yr^{-1}$ of GRB and the assumption of equal injection of kinetic energy into electrons (ultimately observed as photons by synchrotron radiation and, possibly, inverse Compton scattering) and protons by the initial fireball, we calculate a proton flux
\begin{equation}
 E_p \Phi_p = {c\over4\pi} (t_H \dot E) = 2.2\times10^{-10}\rm\, TeV\ (cm^2\, s\ sr)^{-1} \,. 
\end{equation}
Here we assumed injection over a Hubble time $t_H$ of $10^{10}$\,years.

\item From the observed spectrum of the, presumably, extra-galactic cosmic rays beyond the ankle, just above $10^6$\,TeV energy, in the spectrum:
\begin{equation}
E_{\rm CR} \Phi_{\rm CR} = \int_{10^6\rm\ TeV} dE \left[E {dN_{\rm CR}\over dE} \right] \cong 1.7\times 10^{-10}\rm\, TeV\ (cm^2\, s\ sr)^{-1} \,.
\end{equation}
We fitted the spectrum as $E^{-2.7}$ beyond the ankle to obtain this result. The near equality of the GRB and the cosmic ray flux beyond the ankle supports the speculation that GRB are the source of extra-galactic cosmic rays.

\item A final estimate is based on the luminosity of TeV $\gamma$-rays emitted by blazars. Taking the Markarian\,421 flux observed by the Whipple collaboration, and the EGRET luminosity function of 130 sources per steradian:
\begin{equation}
(130\rm\,sr^{-1}) (5\times10^{-10}\,cm^{-2}\, s^{-1} \, TeV) = 6\times10^{-8}\, TeV\ (cm^2\, s\ sr)^{-1} \,.
\end{equation}
This is somewhat less than the observed diffuse $\gamma$-ray flux and over an order of magnitude larger than the proton flux, consistent with the expected hierarchy of photons and protons in a beam dump. This result raises the alternative possibility that AGN are the sources of the highest energy cosmic rays.

\end{enumerate}

From this compilation, it is not unreasonable to assign a luminosity of  $2\times10^{-10}\rm\, TeV\ (cm^2\, s\ sr)^{-1}$ to the source of the highest energy cosmic rays. This corresponds to an injection rate of several times $\dot E = 10^{44}\rm\, erg \, Mpc^{-3}\, yr^{-1}$ and a density of extragalactic cosmic rays of roughly $10^{19}\rm\, erg \, cm^{-3}$. This is very much in line with estimates made elsewhere\cite{taormina}. It is important to keep in mind that estimates of neutrino fluxes using this input represent a {\bf lower limit} because of the absorption of the cosmic rays and TeV gamma rays in the interstellar medium and, possibly, in the source itself.

\subsection{$\nu$'s from AGN}

AGN are the brightest sources in the Universe; some are so far away that they are messengers from the earliest of times. Their engines must not only be powerful, but also extremely compact because their luminosities are observed to flare by over an order of magnitude over time periods as short as a day. Only sites in the vicinity of black holes which are a billion times more massive than our sun, will do. It is anticipated that beams accelerated near the black hole are dumped on the ambient matter in the active galaxy, mostly thermal photons with densities of 10$^{14}$/cm$^3$. The electromagnetic spectrum at all wavelengths, from radio waves to TeV gamma rays, is produced in the interactions of the accelerated particles with the magnetic fields and ambient photons in the galaxy. In most models the highest energy photons are produced by Compton scattering of accelerated electrons on thermal UV photons which are scattered up from 10\,eV to TeV energy\cite{dermer}. The energetic gamma rays will subsequently lose energy by electron pair production in photon-photon interactions in the radiation field of the jet or the galactic disk. An electromagnetic cascade is thus initiated which, via pair production on the magnetic field and photon-photon interactions, determines the emerging gamma-ray spectrum at lower energies. The lower energy photons, observed by conventional astronomical techniques, are, as a result of the cascade process, several generations removed from the primary high energy beams. 

High energy gamma-ray emission (MeV--GeV) has been observed from at least 40 active galaxies by the EGRET instrument on the Compton Gamma Ray Observatory\cite{EGRET}. Most, if not all, are ``blazars". They are AGN viewed from a position illuminated by the cone of a relativistic jet. Of the four TeV gamma-ray emitters conclusively identified by the air Cherenkov technique, two are extra-galactic and are also nearby blazars. The data therefore strongly suggests that the highest energy photons originate in jets beamed at the observer. The cartoon of an AGN, shown in Fig.\,4, displays its most prominent features: an accretion disk of stars and gas falling into the spinning black hole as well as a pair of jets aligned with the rotation axis. Several of the sources observed by EGRET have shown variability, by a factor of 2 or so over a time scale of several days. Time variability is more spectacular at higher energies. On May 7, 1996 the Whipple telescope observed an increase of the TeV-emission from the blazar Markarian 421 by a factor 2 in 1 hour reaching eventually a value 50 times larger than the steady flux. At this point the telescope registered 6 times more photons from the Markarian blazar than from the Crab supernova remnant despite its larger distance (a factor $10^5$). Recently, even more spectacular bursts have been detected from Markarian 501\cite{weekes}.

\begin{figure}[t]
\centering\leavevmode
\epsfxsize=2.5in\epsffile{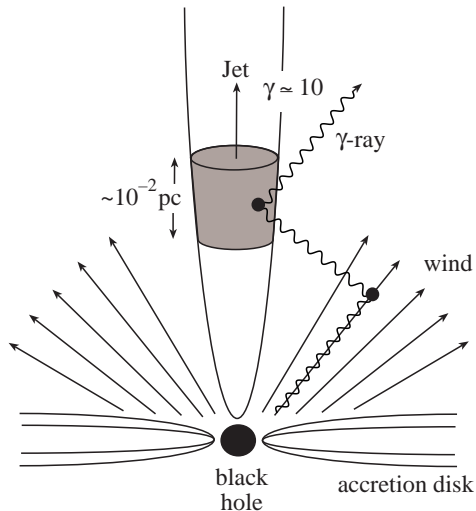}

\caption{Possible blueprint for the production of high energy photons and neutrinos near the super-massive black hole powering an AGN. Particles, electrons and protons(?), accelerated in sheets or blobs moving along the jet, interact with photons radiated by the accretion disk or produced by the interaction of the accelerated particles with the magnetic field of the jet.}
\end{figure}

Pion photoproduction may play a central role in blazar jets. If protons are accelerated along with electrons to PeV--EeV energy, they will produce high energy photons by photoproduction of neutral pions on the ubiquitous UV thermal background. Some have suggested that the accelerated protons initiate a cascade which also dictates the features of the spectrum at lower energy. From a theorist's point of view the proton blazar has attractive features. Protons, unlike electrons, efficiently transfer energy in the presence of the magnetic field in the jet. They provide a ``natural'' mechanism for the energy transfer from the central engine over distances as large as 1\,parsec as well as for the observed heating of the dusty disk over distances of several hundred parsecs.

Although the relative merits of the electron and proton blazar are hotly debated,  it is more relevant that the issues can be settled experimentally. The proton blazar is a source of high energy protons and neutrinos, not just gamma rays. Also, its high energy photon spectrum may exceed the TeV cutoff which is an unavoidable feature of the electron blazar. The opportunities for high energy neutrino astronomy are wonderfully obvious\cite{pr}.

Weakly interacting neutrinos can, unlike high energy gamma-rays and high energy cosmic rays, reach us from more distant and much more powerful AGN. Neutrino astronomers anticipate that the high energy neutrino sky will glow uniformly with bright active galaxies outshining our Milky Way. The results may be even more spectacular. As is the case in man-made beam dumps, photons from celestial accelerators may be absorbed in the dump. The most spectacular sources may therefore have no counterpart in high energy photons. 

Confronted with the challenge to explain a relatively flat multi-wavelength photon emission spectrum reaching TeV energies which is radiated in bursts of a duration of less than one day, models have converged on the blazar blueprint shown in Fig.~4. Particles are accelerated by Fermi shocks in blobs of matter travelling along the jet with a bulk Lorentz factor of $\gamma \sim 10$ and higher. This factor combines the effects of special relativity and the Doppler effect of the moving source; it is also referred to as the Doppler factor. In order to accommodate bursts lasting a day, or less, in the observer's frame, the size of the blob must be of order $\gamma c \Delta t \sim 10^{-2}$\,parsecs or less. The blobs are actually more like sheets, thinner than the jet's size of roughly 1\,parsec. The observed radiation at all wavelengths is produced by the interaction of the accelerated particles in the blob and the ambient radiation in the AGN, which has a significant component concentrated in the so-called ``UV-bump\rlap".

In the following estimate of the neutrino flux from a proton blazar, primes will refer to a frame attached to the blob, which is moving with a Lorentz factor $\gamma$ relative to the observer. In general, the transformation between blob and observer frame is $R' = \gamma R$ and $E' = { 1\over\gamma}E$ for distances and energies, respectively. For a burst of 15\,min duration, the strongest variability observed in TeV emission, the size of the accelerator is only
\begin{equation}
R' = \gamma c \Delta t  \sim 10^{-4} \mbox{ to } 10^{-3}\rm\, pc
\end{equation}
for $\gamma = 10$--$10^2$. So, the jet consists of relatively small structures with short lifetime. High energy emission is associated with the periodic formation of these blobs.

Shocked protons in the blob will photoproduce pions on the photons whose properties are known from the observed multi-wavelength emission. From the observed photon luminosity $L_\gamma$ we deduce the energy density of photons in the shocked region: 
\begin{equation}
U'_\gamma = {L'_\gamma \Delta t\over {4\over3}\pi R'^3} = {L_\gamma \Delta t\over\gamma} \;{1\over {4\over3}\pi (\gamma c\Delta t)^3} = {3\over 4\pi c^3} \; {L_\gamma\over\gamma^4 \Delta t^2} \,.
\end{equation}
(Geometrical factors of order unity will be ignored throughout.)
The dominant photon density is at UV wavelengths, the UV bump. We will assume that a luminosity ${\cal L}_\gamma$ of $10^{45}\rm\,erg\,s^{-1}$ is emitted in photons with energy $E_\gamma = 10$\,eV. Luminosities larger by one order of magnitude have actually been observed. The number density of photons in the shocked region is
\begin{equation}
N'_\gamma = {U'_\gamma\over E'_\gamma} = \gamma {U'_\gamma\over E_\gamma} 
= {3\over 4\pi c^3}\; {L_\gamma \over E_\gamma}\; {1\over\gamma^3 \Delta t^2} \sim 6.8\times10^{14}\mbox{ to }6.8\times10^{11}\rm\, cm^{-3} \,.
\end{equation}
From now on the range of numerical values will refer to $\gamma = 10$--$10^2$, in that order. With such high density the blob is totally opaque to photons with 10\,TeV energy and above. Because photons with such energies have indeed been observed, one must essentially require that 10\,TeV $\gamma$'s are below the $\gamma\gamma\to e^+e^-$ threshold in the blob, i.e., 
\begin{eqnarray}
E_{\rm th} &=& \gamma E'_{\gamma\rm\, th} \agt 10\rm\ TeV,\\
\noalign{\hbox{or}}
E_{\rm th} &>& {m_e^2\over E_\gamma} \gamma^2 > 10\rm\ TeV,\\
\noalign{\hbox{or}}
\gamma &\agt& 10 \,.
\end{eqnarray}
Protheroe et al.\ find $\gamma\agt30$\cite{weekes}. So, we take $10<\gamma<10^2$.

The accelerated protons in the blob will produce pions, predominantly at the $\Delta$-resonance, in interactions with the UV photons. The proton energy for resonant pion production is
\begin{eqnarray}
E'_p &=& {m_{\Delta}^2 - m_p^2\over 4} \; {1\over E'_\gamma}\\
\noalign{\hbox{or}}
E_p &=& {m_\Delta^2 - m_p^2\over 4E_\gamma} \; \gamma^2\\
E_p &=& {1.6\times 10^{17}\rm\,eV\over E_\gamma} \; \gamma^2\\
&=& 1.6\times10^{18}\mbox{ to }1.6\times10^{20}\rm\,eV\,.
\end{eqnarray}
The jet (hopefully) accelerates protons to this energy, and will definitely do so in models where blazars are the sources of the highest energy cosmic rays.

The secondary $\nu_\mu$ have energy
\begin{equation}
E_\nu = {1\over4} \left<x_{p\to\pi}\right> E_p = 7.9\times10^{16}
\mbox{ to }7.9\times10^{18}\rm\ eV
\end{equation}
for $\left<x_{p\to\pi}\right>\simeq 0.2$, the fraction of energy transferred, on average, from the proton to the secondary pion produced via the $\Delta$-resonance. The 1/4 is because each lepton in the decay $\pi\to \mu\nu_\mu\to e\nu_e\nu_\mu\bar\nu_\mu$ carries roughly equal energy.

The fraction of energy $f_\pi$ lost by protons to pion production when travelling a distance $R'$ through a photon field of density $N'_\gamma$ is
\begin{equation}
f_\pi =  {R'\over\lambda_{p\gamma}} = R' N'_\gamma \sigma_{p\gamma\to\Delta} \left< x_{p\to\pi}\right>
\end{equation}
where $\lambda_{p\gamma}$ is the proton interaction length, with $\sigma_{p\gamma\to\Delta\to n\pi^+} \simeq 10^{-28}\rm\,cm^2$. We obtain
\begin{equation}
f_\pi = 3.8\mbox{--}0.038 \mbox{ for } \gamma = 10\mbox{--}10^2 \,.
\end{equation}

For a total injection rate in high-energy protons $\dot E$, the total energy in $\nu$'s is ${1\over2}f_\pi t_H \dot E$, where $t_H =10$\,Gyr is the Hubble time. The factor 1/2 accounts for the fact that 1/2 of the energy in charged pions is transferred to $\nu_\mu + \bar\nu_\mu$, see above. The neutrino flux is
\begin{equation}
\Phi_\nu = {c\over 4\pi} {\left({1\over2} f_\pi t_H \dot E\right)\over E_\nu} e^{f\pi} \,.
\end{equation}
The last factor corrects for the absorption of the protons in the source, i.e., the observed proton flux is a fraction $e^{-f\pi}$ of the source flux which photoproduces pions. We can write this as
\begin{equation}
\Phi_\nu = {1\over E_\nu} {1\over2} f_\pi e^{f\pi} (E_p \Phi_p)\,,
\end{equation}
For $E_p \Phi_p =2\times10^{-10}\rm\ TeV\ (cm^2\, s\ sr)^{-1}$ we obtain
\begin{equation}
\Phi_\nu = 8\times10^5\mbox{ to }2\rm\ (km^2\, yr)^{-1} 
\end{equation}
over $4\pi$ steradians. (Neutrino telescopes are background free for such high energy events and should be able to identify neutrinos at all zenith angles.)

A detailed discussion of how to build high energy neutrino telescopes will be presented further on. For calculational purposes it is sufficient to know that, in order to be detected, i)~a $\nu_\mu$ neutrino must interact in the water or ice near the detector, and ii)~the secondary muon must have a sufficient range to reach the detector.
The detection probability is thus easily computed from the requirement that the neutrino has to interact within a distance of the detector which is shorter than the range of the muon it produces. Therefore,
\begin{equation}
P_{\nu\to\mu} \simeq {R_\mu\over \lambda_{\rm int}} \simeq A E_{\nu}^n \,,
\label{detect-prob}
\end{equation}
where $R_{\mu}$ is the muon range and $\lambda_{\rm int}$ the neutrino interaction length. For energies below 1\,TeV, where both the range and cross section depend linearly on energy, $n=2$. At TeV and PeV energies $n=0.8$ and $A=10^{-6}$, with $E$ in TeV units. For EeV energies $n=0.47$, $A =10^{-2}$ with $E$ in EeV\cite{pr}.

The observed neutrino event rate in a detector is
\begin{equation}
N_{\rm events} = \Phi_\nu P_{\nu\to\mu},
\end{equation}
with
\begin{equation}
P_{\nu\to\mu} \cong 10^{-2} E_{\nu, \rm EeV}^{0.4} \,,
\end{equation}
where $E_\nu$ is expressed in EeV. Therefore
\begin{equation}
N_{\rm events} = (3\times10^3\mbox{ to }5\times 10^{-2})\rm\ km^{-2}\, yr^{-1} = 10^{1\pm2}\rm\, km^{-2}\, yr^{-1}
\end{equation}
for $\gamma = 10\mbox{--}10^2$. This estimate brackets the range of $\gamma$ factors considered. Remember however that the relevant luminosities for protons (scaled to the high energy cosmic rays) and the luminosity of the UV target photons are themselves uncertain.

In summary, for the intermediate value for $\gamma$:
\begin{eqnarray}
E_\nu &=& 7.8\times10^{16}{\rm\,eV} \left(\gamma\over 30\right) \left(\Delta t\over 15\rm\ min\right)\,,\\
f_\pi &=& 0.4 \left(30\over\gamma\right)^2 \left(15{\rm\ min}\over \Delta t\right) \left( {\cal L}_\gamma\over 10^{45}\rm\,erg\ s^{-1}\right)\,,\\
E_\nu &=& 7\times10^5 \left(\gamma\over30\right)^2 \rm\ TeV\,,\\
N_{\rm events} &=& \left(3{\rm\ km^{-2}\,yr^{-1}}\right) \left({f_\pi\over0.4} \, e^{(f_\pi - 0.4)}\right)\nonumber\\
&& {}\times \left(E_{\rm CR} \Phi_{\rm CR}\over 2\times10^{-10}
\rm\ TeV\, cm^{-2}\,s\ sr^{-1}\right) \left(750{\rm\ PeV}\over E_\nu\right)^{0.6}\,.
\end{eqnarray}
Because of further absorption effects on the input proton flux, e.g. in the CMBR, this result should be interpreted as a lower limit.

\subsubsection{The Role of Absorption: Hidden Sources}

The large uncertainty in the calculation of the neutrino flux from AGN is predominantly associated with the boost factor $\gamma$. The reason for this is clear. The target density of photons in the accelerator is determined by i) the photon luminosity, which is directly observed, and ii) the size of the target which is limited by the short duration of the high energy blazar signals. Large boost factors reduce the photon density of the target because they reduce energy and expand the target size in the accelerator frame. A large $\gamma$-factor will render a source transparent to high energy photons and protons despite the high luminosity of photons and the short duration of the burst. Assuming $\gamma=1$, we would conclude instead that the source is completely opaque to high energy photons and protons. It is a ``hidden" source, with reduced or extinct emission of high energy particles, but abundant neutrino production by protons on the high density photon target. The TeV Markarian sources require very large $\gamma$-factors. In their absence the sources would be opaque at TeV energy; see Eq.\,(18). Their neutrino emission is expected to be very low, in the lower range of our prediction. On the other hand, nature presumably made blazars with a distribution of boost factors; observed boost factors are typically less than 10. While uninteresting for high energy gamma ray astronomy, they have the potential to be powerful neutrino emitters, with fluxes near the upper range of our predictions.
   
\subsection{$\nu$'s from GRB}

Recently, GRB may have become the best motivated source for high energy neutrinos. Their neutrino flux can be calculated in a relatively model-independent way.  Although neutrino emission may be less copious and less energetic than from AGN, the
predicted fluxes can probably be bracketed with more confidence.

In GRB a fraction of a solar mass of energy ($\sim10^{53}$\,ergs) is released over a time scale of order 1 second into photons with a very hard spectrum. It has been suggested that, although their ultimate origin is a matter of speculation, the same cataclysmic events also produce the highest energy cosmic rays. This association is reinforced by more than the phenomenal energy and luminosity:

\begin{itemize}

\item
both GRB and the highest energy cosmic rays are produced in cosmological
sources, {\it i.e.}, and, as previously discussed,

\item
the average rate at which energy is injected into the Universe as gamma rays from GRB
is similar to the rate at which energy must be injected in the
highest energy cosmic rays in order to produce the observed cosmic ray
flux beyond the ``ankle'' in the spectrum at $10^7$\,TeV.

\end{itemize}

There is increasing observational support for a model where an initial
event involving neutron stars or black holes deposits a solar mass of
energy into a radius of order 100\,km. Such a state is opaque
to light. The observed gamma ray display is the result of a
relativistic shock with $\gamma = 10^2\mbox{--}10^3$ which expands the original fireball by a factor $10^6$ over 1\,second. Gamma rays are produced by synchrotron radiation by
relativistic electrons accelerated in the shock, possibly followed by
inverse-Compton scattering. The association of cosmic rays with GRB
obviously requires that kinetic energy in the shock is converted into the
acceleration of protons as well as electrons. It is assumed that the
efficiency with which kinetic energy is converted to accelerated protons is
comparable to that for electrons.  The production of high-energy neutrinos is a feature of the fireball model because the protons will photoproduce pions and, therefore, neutrinos on the gamma rays in the burst. We have a beam dump configuration where both the beam and target are constrained by observation: of the cosmic ray beam and of the photon fluxes at Earth, respectively.

\begin{figure}[h]
\centering\leavevmode
\epsfxsize=3.5in\epsffile{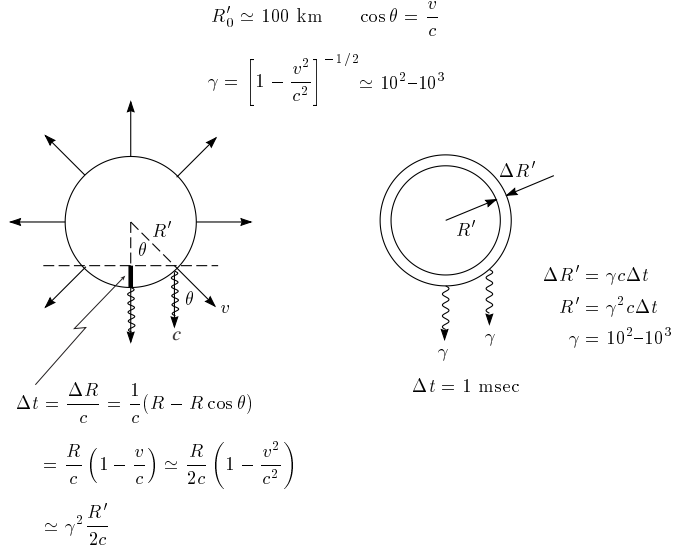}

\caption{Kinematics of GRB.}
\end{figure}

Simple relativistic kinematics (see Fig.\,5) relates the radius and width $R', \Delta R'$ to the observed duration of the photon burst $c\Delta t$ 
\begin{eqnarray}
R' &=& \gamma^2 (c\Delta t)\\
\Delta R' &=& \gamma c\Delta t
\end{eqnarray}
The calculation of the neutrino flux follows the same path as that for AGN. From the observed GRB luminosity $L_\gamma$ we compute the photon density in the shell:
\begin{equation}
U'_\gamma = { \left( L_\gamma \Delta t / \gamma\right) \over 4\pi
R'^2 \Delta R' } = {L_\gamma\over 4\pi R'^2 c\gamma^2} 
\end{equation}
The pion production by shocked protons in this photon field is, as before, calculated from the interaction length
\begin{eqnarray}
&& {1\over\lambda_{p\gamma}} = N_\gamma \sigma_\Delta \left< x_{p\to\pi}\right> =
{U'_\gamma\over E'_\gamma} \sigma_\Delta \left< x_{p\to\pi}\right> \qquad
\left(E'_\gamma = {1\over\gamma}E_\gamma\right) \\
\end{eqnarray}
As before, $\sigma_\Delta$ is the cross section for $p\gamma\to\Delta\to n\pi^+$ and $\left< x_{p\to\pi}\right> \simeq 0.2$. The fraction of energy going into $\pi$-production is
\begin{eqnarray}
    f_\pi &\cong& {\Delta R'\over\lambda_{p\gamma}}\\
    f_\pi &\simeq& {L_\gamma\over E_\gamma} {1\over\gamma^4\Delta t}
    {\sigma_\Delta\left< x_{p\to\pi}\right>\over 4\pi c^2}\\
    f_\pi &\simeq& 0.14 \left\{ L_\gamma\over 10^{51}{\rm\,ergs}^{-1} \right\}
    \left\{ 1{\rm\ MeV}\over E_\gamma\right\}
    \left\{300\over\gamma\right\}^4 \left\{ 1{\rm\ msec}\over\Delta t\right\}
\nonumber\\
   && \hskip1.5em\times \left\{ \sigma_\Delta\over 10^{-28}{\rm\,cm^2} \right\}
     \left\{ \left< x_{p\to\pi}\right> \over 0.2\right\}
\end{eqnarray}
The relevant photon energy in the problem is 1\,MeV, the energy where the typical GRB spectrum exhibits a break. The number of higher energy photons is suppressed by the spectrum, and lower energy photons are less efficient at producing pions. Given the large uncertainties associated with the astrophysics, it is an adequate approximation to neglect the explicit integration over the GRB photon spectrum. The proton energy for production of  pions via the $\Delta$-resonance is   
\begin{eqnarray}
E'_p &=& { m_\Delta^2 - m_p^2\over 4 E'_\gamma}
 \end{eqnarray}
Therefore,
\begin{eqnarray}
E_p &=& 1.4\times10^{16}\rm\, eV \left(\gamma\over300\right)^2
\left(1\ MeV\over E_\gamma\right)\\
E_\nu &=& {1\over4} \left< x_{p\to\pi}\right> E_p \simeq 7\times10^{14} \rm\, eV
\end{eqnarray}
We are now ready to calculate the neutrino flux:
\begin{equation}
\phi_\nu = {c\over4\pi} {U'_\nu\over E'_\nu} = {c\over4\pi}
{U_\nu\over E_\nu} = {c\over 4\pi} {1\over E_\nu} \left\{ {1\over2} f_\pi
t_H \dot E\right\}
\end{equation}
where, as before, the factor 1/2 accounts for the fact that only 1/2 of the energy in charged pions is transferred to $\nu_\mu + \bar\nu_\mu$. As before, $\dot E$ is the injection rate in cosmic rays beyond the ankle (${\sim} 4\times
10^{44}\rm\,erg\,Mpc^{-3}\,yr^{-1}$) and $t_H$ is the Hubble time of ${\sim}10^{10}$\,Gyr. Numerically,
\begin{eqnarray}
\phi_\nu &=& 2\times10^{-14}{\rm\, cm^{-2}\, s^{-1}\, sr^{-1}}
     \left\{ 7\times10^{14}\rm\, eV\over E_\nu \right\} \left\{ f_\pi\over
0.125\right\} \left\{t_H\over 10\rm\ Gyr\right\}\nonumber\\
 && \hspace{1.75in} {}\times \left\{ \dot E\over4\times10^{44} \rm\, erg\,Mpc^{-3}\,yr^{-1} \right\}
\end{eqnarray}
The observed muon rate is
\begin{eqnarray}
N_{\rm events} &=& \int_{E_{th}}^{E_\nu^{\rm max}} \Phi_\nu
P_{\nu\to\mu} {dE_\nu\over E_\nu},\\
\end{eqnarray}
where $P_{\nu\to\mu} \simeq 1.7\times10^{-6}E_\nu^{0.8}$\,(TeV) for TeV energy. Therefore
\begin{eqnarray}
N_{\rm events} &\cong& 26\rm\ km^{-2}\,yr^{-1} \left\{ E_\nu\over
7\times10^{14}\,eV\right\}^{-0.2} \left\{ \Delta\theta\over4\pi\right\}
\end{eqnarray}

The result is insensitive to beaming. Beaming yields more energy per burst, but less bursts are actually observed. The predicted rate is also insensitive to the neutrino energy $E_\nu$ because higher average energy yields less $\nu$'s, but more are detected. Both effects are approximately linear.

There is also the possibility that high-energy gamma rays and neutrinos are produced when the shock expands further into the interstellar medium. This mechanism has been invoked as the origin of the delayed high energy gamma rays. The fluxes are produced over seconds, possibly longer. It is easy to adapt the previous calculation to the external shock. Following, for instance, Bottcher et al., the time scale is changed from milliseconds to seconds and the break in the spectrum from 1 to 0.1\,MeV, we find that $f_{\pi}$ is reduced by two orders of magnitude. In the external shocks higher energies can be reached (a factor 10 higher the for Bottcher et al.\ model) and this increases the neutrino detection efficiency. In the end, the observed rates are an order of magnitude smaller, but the inherent ambiguities of the estimates are such that it is difficult to establish with confidence the relative rate in internal and external shocks. Again, the result boosts the argument for neutrino telescopes of kilometer scale.

\section{Large Natural Cherenkov Detectors}

Neutrino telescopes are conventional particle detectors which use natural and clear water and ice as a Cherenkov medium. A three dimensional grid of photomultiplier tubes maps the Cherenkov cone radiated by a muon of neutrino origin. Nanosecond timing provides degree resolution of the muon track which is, at least for high energy neutrinos, aligned with the neutrino direction. The detectors are shielded from the flux of cosmic ray muons by a kilometer, or more, of water and ice. Yet, identifying neutrinos in this down-going muon background is impossible. Cosmic ray muons exceed those of neutrino origin by a factor $10^5$, or more, depending on the depth of the instrument. Only up-going muons made by neutrinos reaching us through the Earth can be successfully detected. The Earth is used as a filter to screen cosmic ray muons which makes neutrino detection possible over the lower hemisphere of the detector.

The probability to detect a TeV neutrino is roughly $10^{-6}$. As previously discussed, this is easily computed from the requirement that, in order to be detected, the neutrino has to interact within a distance of the detector which is shorter than the range of the muon it produces; see Eq.~(\ref{detect-prob}). At PeV energy the cosmic ray flux is of order 1 per m$^{-2}$ per year and the probability to detect a neutrino of this energy is of order 10$^{-3}$. A neutrino flux equal to the cosmic ray flux will therefore yield only a few events per day in a kilometer squared detector. At EeV energy the situation is worse. With a rate of 1 per km\,$^2$ per year and a detection probability of 0.1, one can still detect several events per year in a kilometer squared detector provided the neutrino flux exceeds the proton flux by 2 orders of magnitude or more. For the neutrino flux generated by cosmic rays interacting with CMBR photons and such sources as AGN and topological defects\cite{schramm}, this is indeed the case. All above estimates are conservative and the rates should be higher because the neutrinos escape the source with a flatter energy spectrum than the protons. In summary, at least where cosmic rays are part of the beam dump, their ray flux and the neutrino cross section and muon range define the size of a neutrino telescope. Needless to say that a telescope with kilometer squared effective area represents a neutrino detector of kilometer cubed volume.

\subsection{Baikal and the Mediterranean}

First generation neutrino telescopes, launched by the bold decision of the
DUMAND collaboration over 25 years ago to construct such an instrument, are
designed to reach a relatively large telescope area and detection volume for a
neutrino threshold of 1--100~GeV. This relatively low threshold
permits calibration of the novel instrument on the known flux of atmospheric
neutrinos.  Its architecture is optimized for reconstructing the Cherenkov
light front radiated by an up-going, neutrino-induced muon. Up-going muons are
to be identified in a background of down-going, cosmic ray muons which are
more than $10^5$ times more frequent for a depth of 1$\sim$2 kilometers. The
method is sketched in Fig.\,6.

\begin{figure}[h]
\centering
\hspace{0in}\epsfxsize=2.1in\epsffile{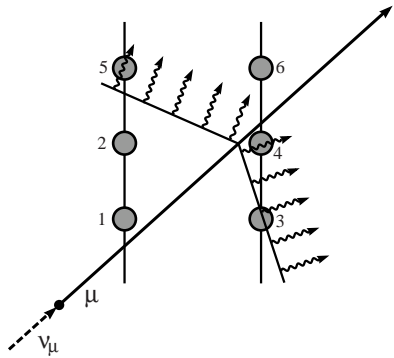}

\caption{The arrival times of the Cherenkov photons in 6 optical modules
determine the direction of the muon track.}
\end{figure}

The ``landscape" of neutrino astronomy is sketched in Table~4. With the termination of the pioneering DUMAND experiment, the efforts in water are, at present, spearheaded by the Baikal experiment\cite{domogatsky}. Operating with 144 optical modules (OM) since April 1997, the {\it NT-200} detector has been completed in April 1998. The Baikal detector is well understood and the first atmospheric neutrinos have been identified; we will discuss this in more detail further on. The Baikal site is competitive with deep oceans although the smaller absorption length of Cherenkov light in lake water requires a somewhat denser spacing of the OMs. This does however result in a lower threshold which is a definite advantage, for instance in WIMP searches. They have shown that their shallow depth of 1 kilometer does not represent a serious drawback. By far the most significant advantage is the site with a seasonal ice cover which allows reliable and inexpensive deployment and repair of detector elements.

In the following years, {\it NT-200} will be operated as a neutrino telescope with an effective area between $10^3 \sim 5\times 10^3$\,m$^2$, depending on the energy. Presumably too small to detect neutrinos from AGN and other extraterrestrial sources, {\it NT-200} will serve as the prototype for a larger telescope. For instance, with 2000 OMs, a threshold of  $10 \sim 20$\,GeV and an effective area of $5\times10^4 \sim 10^5$\,m$^2$, an expanded Baikal telescope would fill the gap between present underground detectors and planned high threshold detectors of cubic kilometer size. Its key advantage would be low threshold.

\begin{table}[t]
\caption{}
\medskip
\centering\leavevmode
\epsfxsize=3in\epsffile{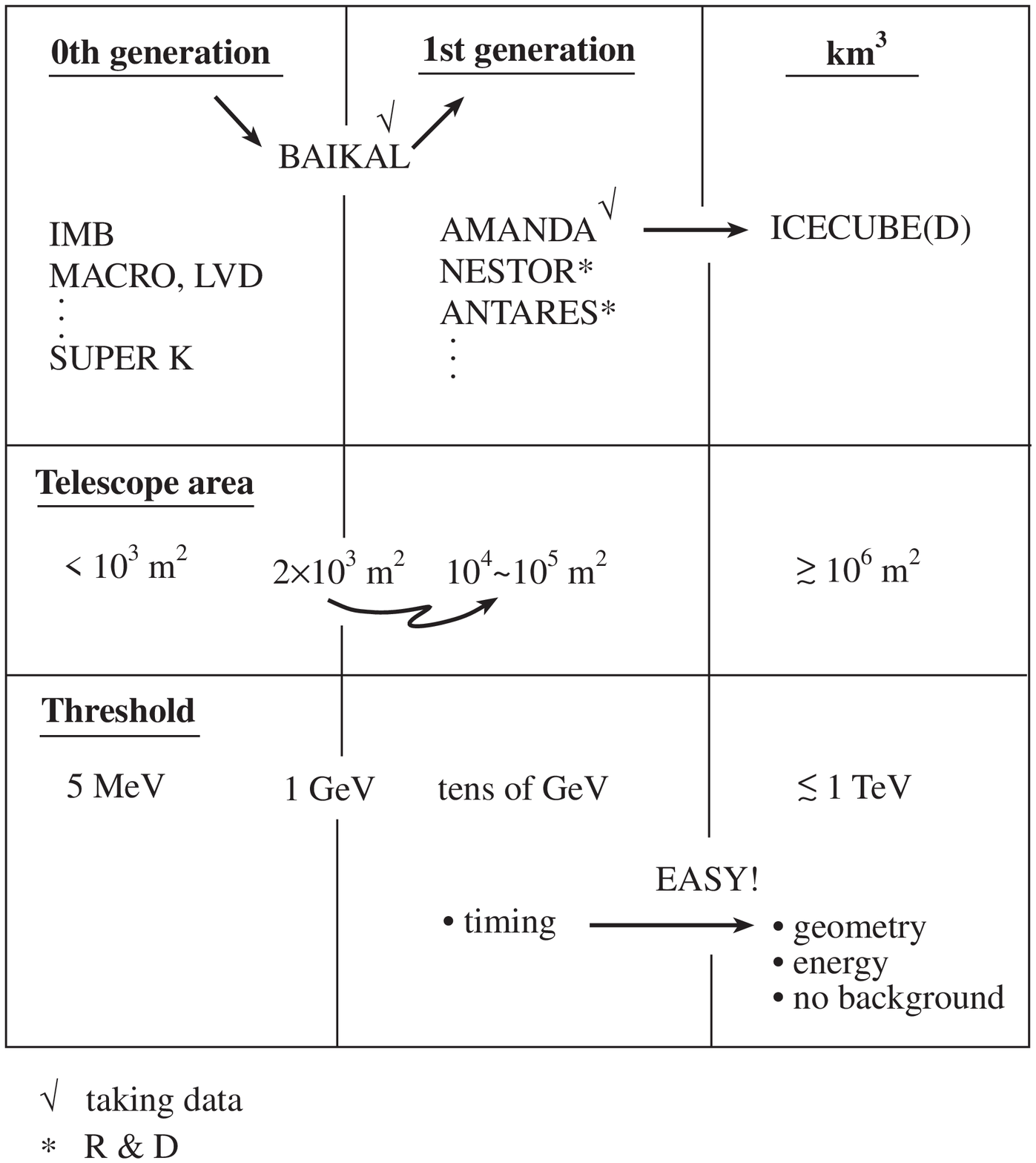}
\end{table}

The Baikal experiment represents a proof of concept for deep ocean projects. These should have the advantage of larger depth and optically superior water. Their challenge is to design a reliable and affordable technology. Several groups are confronting the problem, both NESTOR and Antares are developing rather different detector concepts in the Mediterranean.

The NESTOR collaboration\cite{resvanis}, as part of an ongoing series of technology tests, are testing the umbrella structure which will hold the OMs. They deployed two aluminium ``floors", 34\,m in diameter, to a depth of 2600\,m. Mechanical robustness was demonstrated by towing the structure, submerged below 2000\,m, from shore to the site and back. The detector will consist of 8 six-legged floors separated by 30\,m.

The Antares collaboration\cite{feinstein} is in the process of determining the critical detector parameters at a 2000\,m deep, Mediterranean site off Toulon, France. First results on water quality are very encouraging. They have recently demonstrated their capability of deploying and retrieving a string. A deliberate development effort will lead to the construction of a demonstration project consisting of 3 strings with a total of 200 OMs.

For neutrino astronomy to become a viable science several of these, or other, projects will have to succeed. Astronomy, whether in the optical or in any other wave-band, thrives on a diversity of complementary instruments, not on ``a single best instrument\rlap." When the Soviet government tried out the latter method by creating a national large mirror project, it virtually annihilated the field.

\subsection{First Neutrinos from Baikal}

The Baikal Neutrino Telescope is deployed in Lake Baikal, Siberia, 3.6\,km from shore at a depth of 1.1\,km. An umbrella-like frame holds 8 strings, each instrumented with 24 pairs of 37-cm diameter {\it QUASAR} photomultiplier tubes (PMT). Two PMTs in a pair are switched in coincidence in order to suppress background from bioluminescence and PMT noise.

They have analysed 212 days of data taken in 94-95 with 36 OMs. Upward-going muon candidates were selected from about $10^{8}$ events in which more than 3 pairs of PMTs triggered. After quality cuts and $\chi^2$ fitting of the tracks a sample of 17 up-going events remained. These are not generated by neutrinos passing the Earth below the detector, but by showers from down-going muons originating below the array. In a small detector such events are expected. In 2 events however the light intensity does not decrease from bottom to top, as expected from invisible showering muons below the detector. A detailed analysis \cite{ourneu} yields a fake probability of 2\% for both events.

After the deployment of 96 OMs in the spring of 96, three neutrino candidates have been found in a sample collected over 18 days. This is in agreement with the expected number of approximately 2.3 for neutrinos of atmospheric origin.  One of the events is displayed in Fig.\,7. In this analysis the most effective quality cuts are the traditional $\chi^2$ cut and a cut on the probability of non-reporting channels not to be hit, and reporting channels to be hit ($P_{nohit}$ and $P_{hit}$, respectively). To guarantee a minimum lever arm for track fitting, they were forced to reject events with a projection of the most distant channels on the track smaller than 35 meters. This does, of course, result in a loss of threshold.

\begin{figure}[t]
\centering\leavevmode
\epsfxsize=1.5in\epsffile{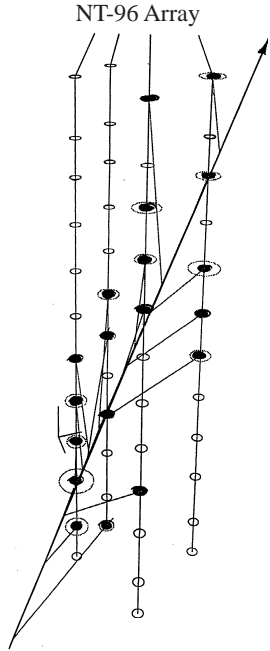}

\caption{Candidate neutrino event from NT-96 in Lake Baikal.}
\end{figure}

\subsection{The AMANDA South Pole Neutrino Detector}

\subsubsection{Status of the AMANDA Project}

Construction of the first-generation AMANDA detector\cite{barwick} was completed in the austral summer 96--97. It consists of 300 optical modules
deployed at a depth of 1500--2000~m; see Fig.\,8. An optical module (OM) consists of an 8~inch photomultiplier tube and nothing else. OM's have only
failed when the ice refreezes, at a rate of less than 3 percent. Detector calibration and analysis of the first year of data is in progress, although data has been taken with 80 calibrated OM's which were deployed one year earlier in order to verify the optical properties
of the ice below 1~km depth (AMANDA-80).

\begin{figure}[t]
\centering
\hspace{0in}\epsfxsize=3.85in%
\epsffile{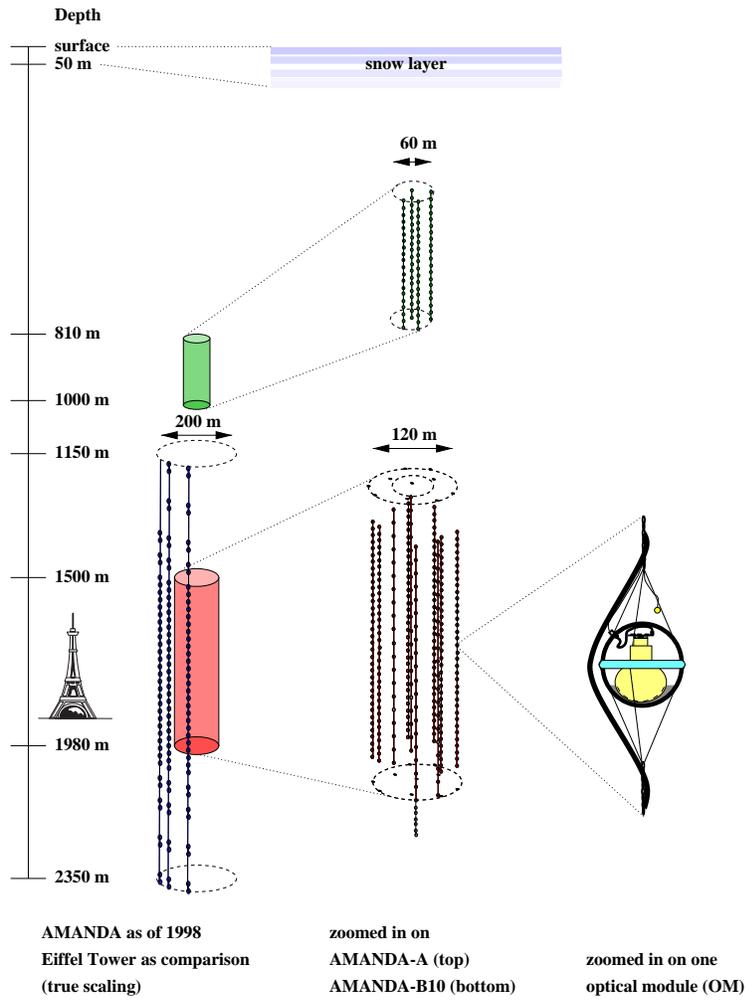}

\caption{The Antarctic Muon And Neutrino Detector Array (AMANDA).}
\end{figure}

As anticipated from transparency measurements performed with the shallow
strings\cite{science} (see Fig.\,8), we found that ice is bubble-free at 1400--1500~meters and below. The performance of the AMANDA detector is encapsulated in the event shown in Fig.\,9. Coincident events between AMANDA-80
and four shallow strings with 80 OM's have been triggered for one year at a
rate of 0.1~Hz. Every 10 seconds a cosmic ray muon is tracked over 1.2 kilometers. The contrast in detector response between the strings near 1 and
2~km depths is dramatic: while the Cherenkov photons diffuse on remnant bubbles in the shallow ice, a straight track with velocity $c$ is registered
in the deeper ice. The optical quality of the deep ice can be assessed by
viewing the OM signals from a single muon triggering 2 strings separated by
77.5~m; see Fig.\,9b. The separation of the photons along the Cherenkov cone is
well over 100~m, yet, despite some evidence of scattering, the speed-of-light
propagation of the track can be readily identified.

\renewcommand{\thefigure}{\arabic{figure}a}
\begin{figure}[t]
\centering
\hspace{0in}\epsfxsize=3.5in\epsffile{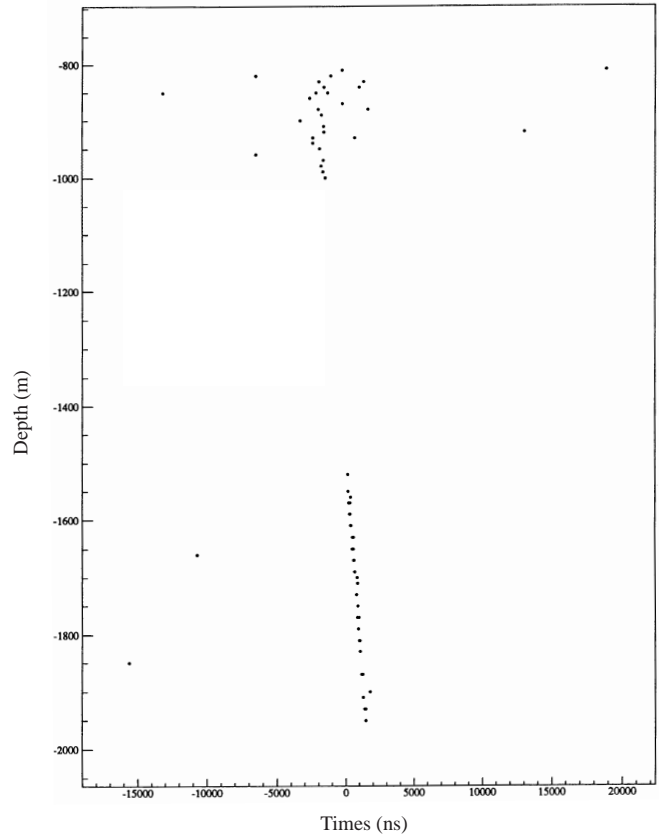}

\caption{Cosmic ray muon track triggered by both shallow and deep AMANDA
OM's. Trigger times of the optical modules are shown as a function of depth.
The diagram shows the diffusion of the track by bubbles above 1~km depth.
Early and late hits, not associated with the track, are photomultiplier noise.}
\end{figure}

\addtocounter{figure}{-1}\renewcommand{\thefigure}{\arabic{figure}b}
\begin{figure}[t]
\centering
\hspace{0in}\epsfxsize=3.5in\epsffile{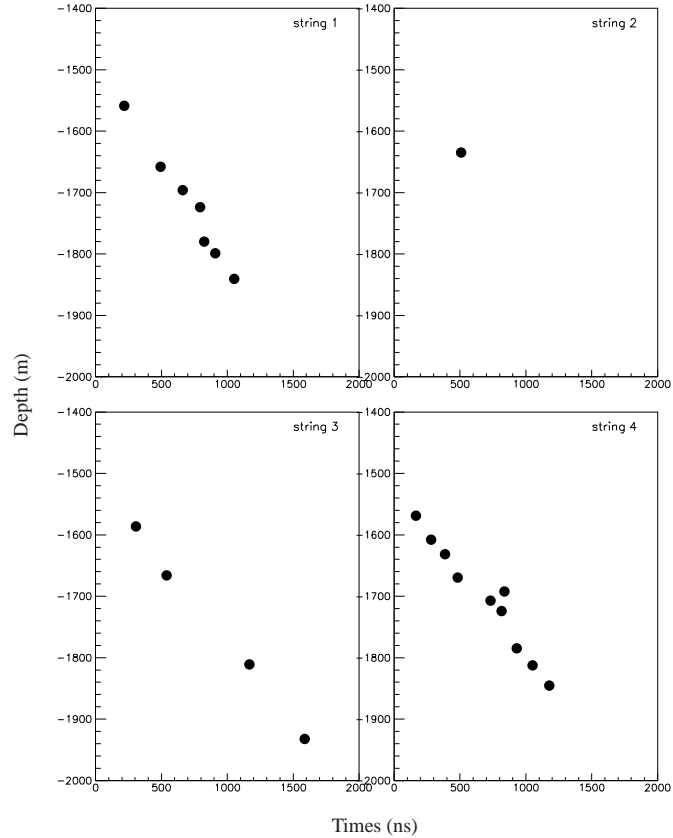}

\caption{Cosmic ray muon track triggered by both shallow and deep AMANDA
OM's. Trigger times are shown separately for each string in the deep detector.
In this event the muon mostly triggers OM's on strings 1 and 2 which are
separated by 77.5~m. }
\end{figure}
\renewcommand{\thefigure}{\arabic{figure}}

The optical properties of the ice are quantified by studying the propagation
in the ice of pulses of laser light of nanosecond duration. The arrival times
of the photons after 20~m and 40~m are shown in Fig.\,10 for the shallow and
deep ice\cite{serap}. The distributions have been normalized to equal areas;
in reality, the probability that a photon travels 70~m in the deep ice is
${\sim}10^7$ times larger. There is no diffusion resulting in loss of information on the geometry of the Cherenkov cone in the deep bubble-free ice. These critical results have been verified by the deployment of nitrogen lasers, pulsed LED's and DC lamps in the deep ice. TV cameras have been lowered to 2400~m.

\begin{figure}[t]
\centering
\hspace{0in}\epsfxsize=4.25in\epsffile{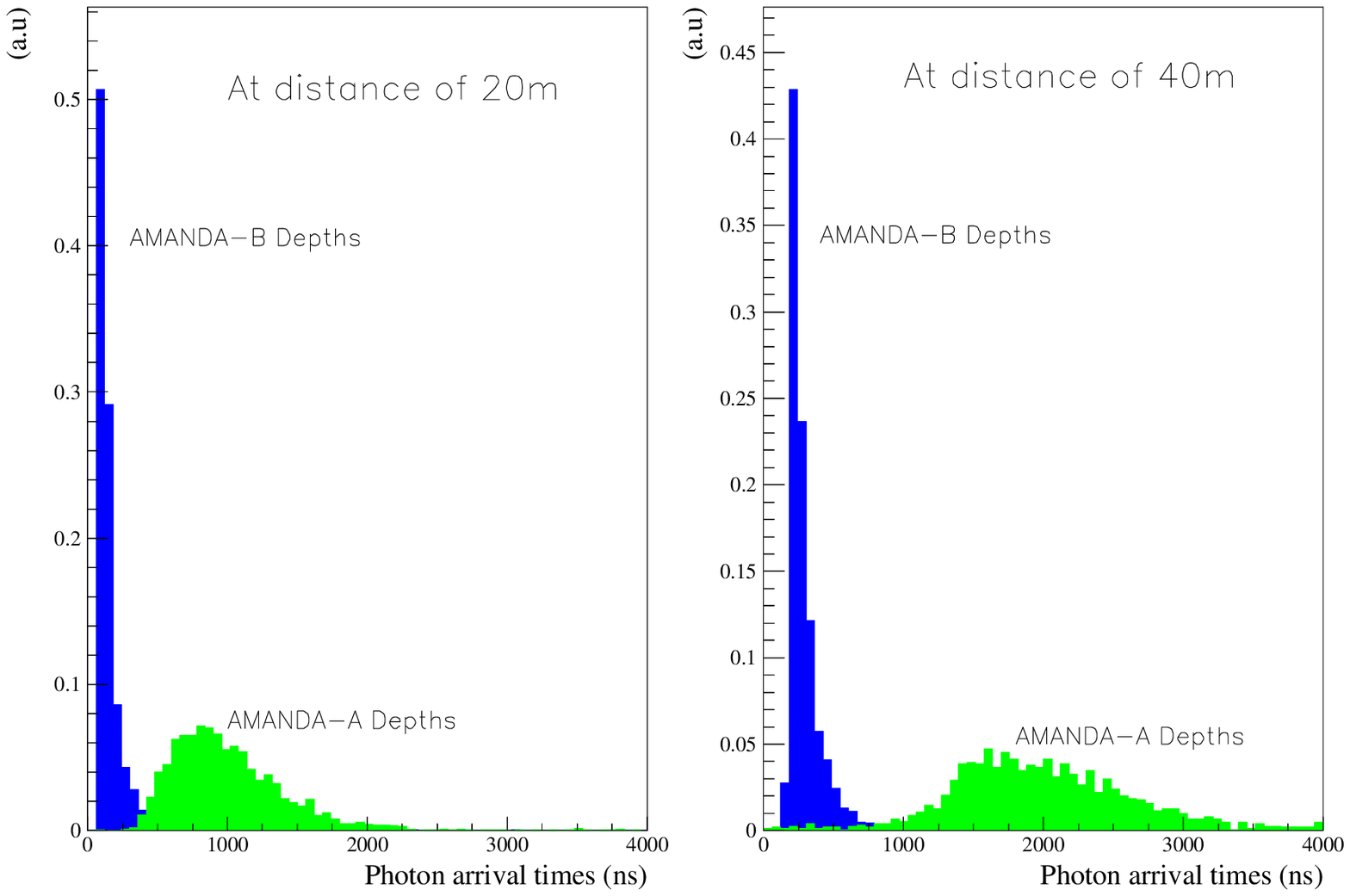}

\caption{Propagation of 510~nm photons indicate bubble-free ice below 1500~m,
in contrast to ice with some remnant bubbles above 1.4~km.}
\end{figure}

\subsubsection{AMANDA: before and after}

The AMANDA detector was antecedently proposed on the premise that inferior
properties of ice as a particle detector with respect to water could be compensated by additional optical modules. The technique was supposed to be a
factor $5 {\sim} 10$ more cost-effective and, therefore, competitive. The
design was based on then current information\cite{dublin}:

\begin{itemize}
\item
the absorption length at 370~nm, the wavelength where photomultipliers are
maximally efficient, had been measured to be 8~m;

\item
the scattering length was unknown;

\item
the AMANDA strategy would have been to use a large number of closely spaced
OM's to overcome the short absorption length. Muon tracks triggering 6 or more
OM's are reconstructed with degree accuracy. Taking data with a simple majority trigger of 6 OM's or more at 100~Hz yields an average effective area
of $10^4$~m$^2$, somewhat smaller for atmospheric neutrinos and significantly
larger for the high energy signals previously discussed.
\end{itemize}

\noindent
The reality is that:
\begin{itemize}
\item
the absorption length is 100~m or more, depending on depth\cite{science};

\item
the scattering length is ${\sim} 25$~m (preliminary, this number represents an average value which may include the combined effects of deep ice
and the refrozen ice disturbed by the hot water drilling);

\item
because of the large absorption length, OM spacings are similar, actually
larger, than those of proposed water detectors. Also, in a trigger 20 OM's report, not 6. Of these more than 5 photons are, on average,
``not scattered\rlap." A precise definition of ``direct" photons will be given
further on. In the end, reconstruction is therefore as before, although additional information can be extracted from scattered photons by minimizing a
likelihood function which matches measured and expected
delays\cite{christopher}.
\end{itemize}

The measured arrival directions of background cosmic ray muon tracks, reconstructed with 5 or more unscattered photons, are confronted with their
known angular distribution in Fig.\,11. There is an additional cut in Fig.\,11
which simply requires that the track, reconstructed from timing information,
actually traces the spatial positions of the OM's in the trigger. The power of
this cut, especially for events distributed over only 4 strings, is very
revealing. It can be shown that, in a kilometer-scale detect geometrical
track reconstruction using only the positions of triggered OM's is sufficient
to achieve degree accuracy in zenith angle. We conclude from Fig.\,11 that the
agreement between data and Monte Carlo simulation is adequate. Less than one
in $10^5$ tracks is misreconstructed as originating below the
detector\cite{serap}. Visual inspection reveals that the remaining misreconstructed tracks are mostly showers, radiated by muons or initiated by
electron neutrinos, misreconstructed as up-going tracks of muon neutrino
origin. At the $10^{-6}$ level of the background, up-going muon tracks can be
identified; see Fig.\,12. Showers can be readily
eliminated on the basis of the additional information on the amplitude of OM
signals. The rate of tracks reconstructed as up-going is consistent with atmospheric neutrino origin of these events; see next section.

\begin{figure}[t]
\centering
\leavevmode\epsfxsize=4in%
\epsffile{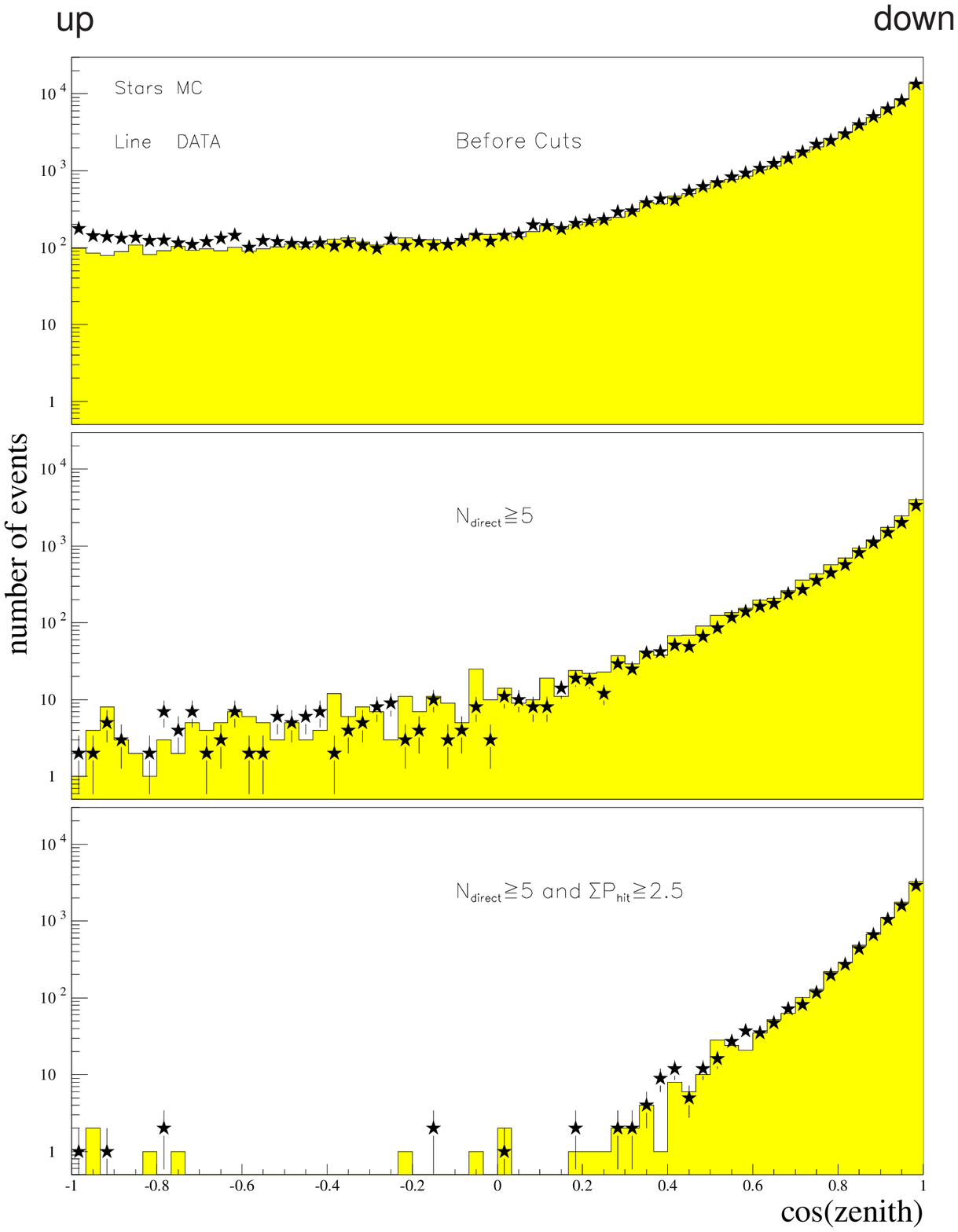}

\caption{Reconstructed zenith angle distribution of muons triggering AMANDA-80: data and Monte Carlo. The relative normalization has not been
adjusted at any level. The plot demonstrates a rejection of cosmic ray muons
at a level of 10$^{-5}$.}
\end{figure}

\begin{figure}[t]
\centering
\hspace{0in}\epsfysize=4.5in\epsffile{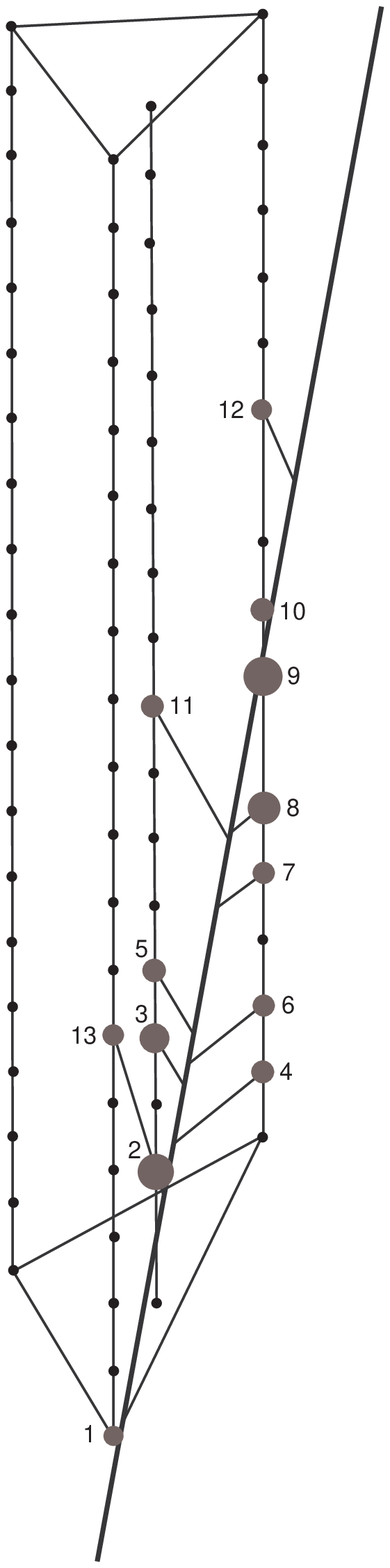}

\caption{A muon reconstructed as up-going in the AMANDA-80 data.
The numbers show the time sequence of triggered OMs, the size of the dots the
relative amplitude of the signal.}
\end{figure}

Monte Carlo simulation, based on this exercise, anticipates that AMANDA-300 is a $10^4$~m$^2$ detector for TeV muons, with 2.5 degrees mean angular resolution per event\cite{christopher}. We have verified the angular resolution of
AMANDA-80 by reconstructing muon tracks registered in coincidence with a
surface air shower array SPASE\cite{miller}. Figure~13 demonstrates that the
zenith angle distribution of the coincident SPASE-AMANDA cosmic ray beam
reconstructed by the surface array is quantitatively reproduced by reconstruction of the muons in AMANDA.

\begin{figure}[t]
\centering\leavevmode
\epsfxsize=3.5in\epsffile{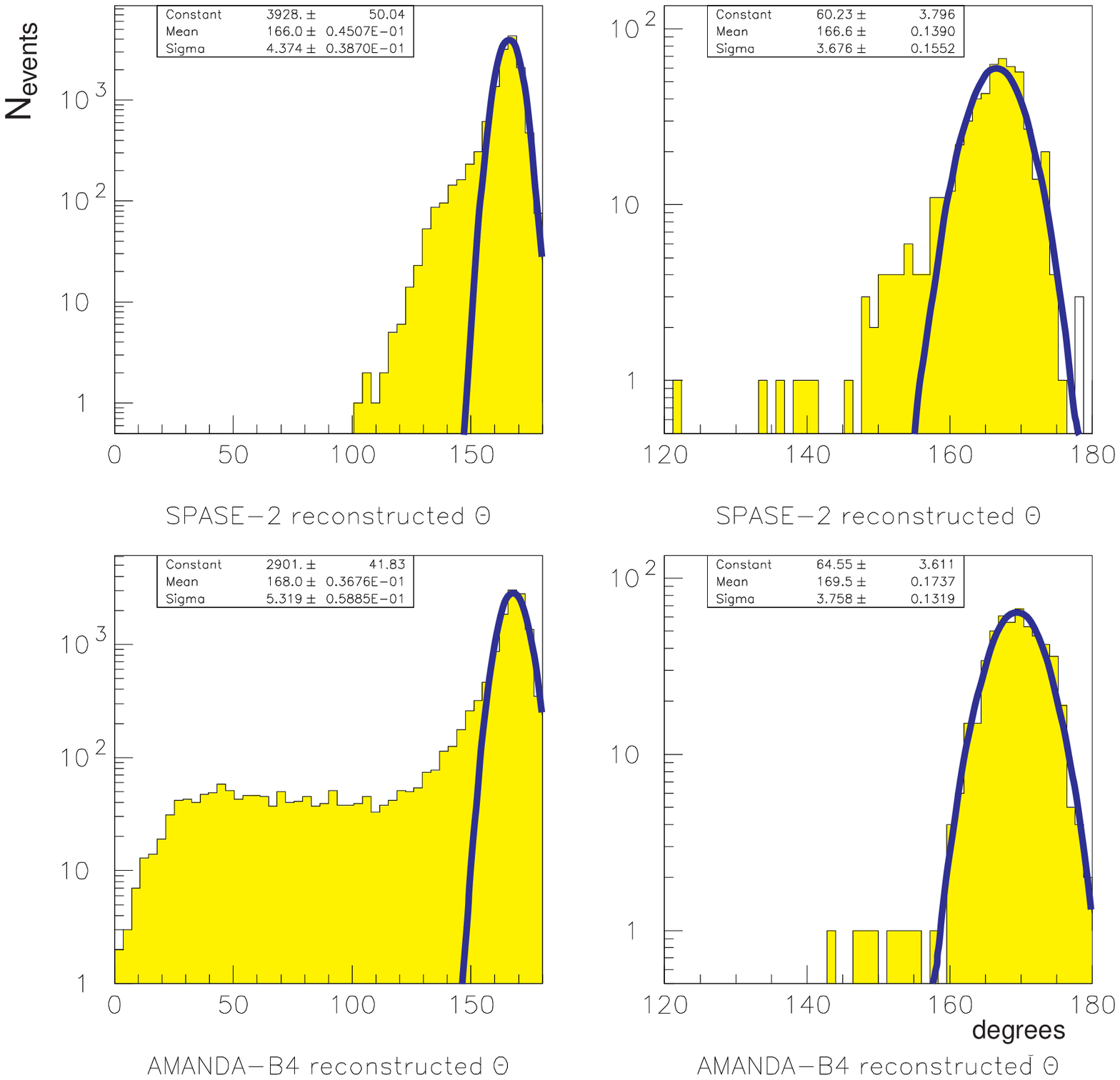}

\caption{Zenith angle distributions of cosmic rays triggering AMANDA and the
surface air shower array SPASE. Reconstruction by AMANDA of underground muons
agrees with the reconstruction of the air shower direction using the scintillator array, and with Monte Carlo simulation. The events are selected requiring signals on 2 or more strings (left), and 5 or more direct photons (right).}
\end{figure}

\section{Neutrinos from the Earth's Center: AMANDA-80 events}

The capability of neutrino telescopes to discover the particles that constitute the dominant, cold component of the dark matter has been previously mentioned. The existence of the weakly interacting massive particles
(WIMPs) in our galactic halo is inferred from observation of their annihilation products. Cold
dark matter particles annihilate into neutrinos; {\it massive} ones will
annihilate into {\it high-energy} neutrinos which can be detected in high-energy neutrino telescopes. This so-called indirect detection is greatly
facilitated by the fact that the Earth and the sun represent dense, nearby
sources of accumulated cold dark matter particles. Galactic WIMPs, scattering
off nuclei in the sun, lose energy. They may fall below escape velocity and be
gravitationally trapped. Trapped WIMPs eventually come to equilibrium and
accumulate near the center of the sun. While the WIMP density builds up, their
annihilation rate into lighter particles increases until equilibrium is achieved where the annihilation rate equals half of the capture rate. The sun
has thus become a reservoir of WIMPs which we expect to annihilate mostly into
heavy quarks and, for the heavier WIMPs, into weak bosons. The leptonic decays of the heavy quark and weak boson annihilation products turn the sun
and Earth into nearby sources of high-energy neutrinos with energies in the
GeV to TeV range. Existing neutrino detectors have already excluded fluxes of neutrinos from the Earth's center of order 1~event per $1000 \rm~m^2$ per year. The
best limits have been obtained by the Baksan experiment\cite{suvorova}. They are already excluding relevant parameter space of supersymmetric models. We will
show that, with data already on tape, the AMANDA detector will have an unmatched discovery reach for WIMP masses in excess of 500~GeV.

We have performed a search\cite{bouchta} for upcoming neutrinos from the
center of the Earth. One should keep in mind that the preliminary results are
obtained with only 80 OMs, incomplete calibration of the optical modules and only 6 months of data. We nevertheless obtain limits near the competitive level of less than 1 event
per $250\rm\,m^2$ per year for WIMP masses in excess of 100~GeV. Increased sensitivity should result from: lower threshold, better calibration (factor of 3), improved angular resolution (factor of $\sim$2), longer exposure and, finally, an effective area larger by over one order of magnitude. Recall that, because the search is limited by atmospheric neutrino
background, sensitivity only grows as the square root of the effective area. First calibration of the full detector is now completed and analysis of the first year of data is in progress. Preliminary results based on the analysis of 1 month of data confirm the performance of the detector derived from the analysis of AMANDA-B4 data. Events reconstructed as going  upwards, like the one shown in Fig.\,14, are found, as expected.

\begin{figure}
\centering\leavevmode
\epsfxsize=4.5in\epsffile{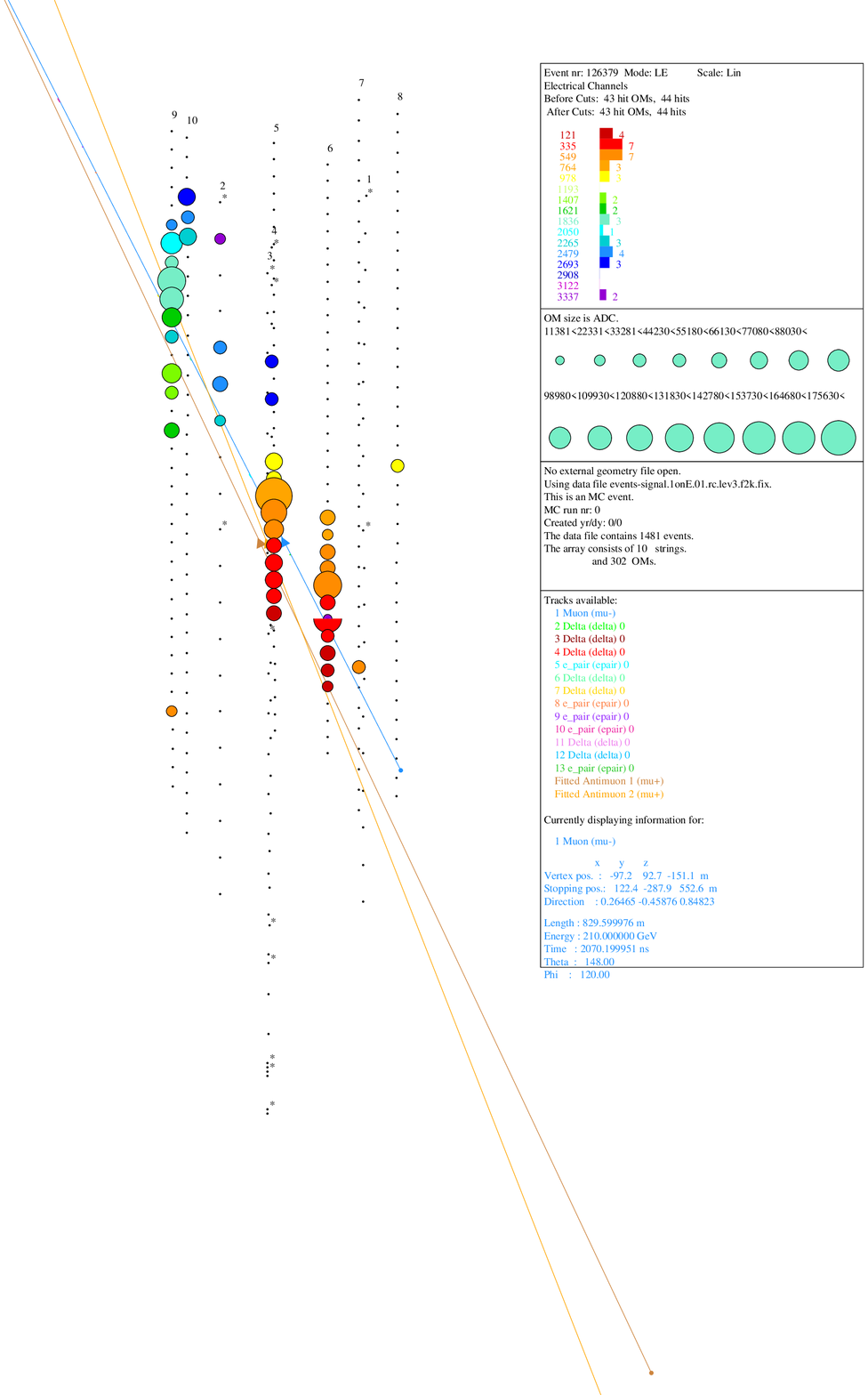}

\caption{}
\end{figure}

We reconstructed 6 months of filtered AMANDA-80 events subject to the conditions that 8 OMs report a signal in a time window of 2 microseconds.
While the detector accumulated data at a rate of about 20~Hz, filtered events
passed cuts\cite{jacobsen} which indicate that time flows upwards through the
detector. In collider experiments this would be referred to as a level 3
trigger. The narrow, long AMANDA-80 detector (which constitutes the 4 inner
strings of AMANDA-300) thus achieves optimal efficiency for muons pointing
back towards the center of the Earth which travel vertically upwards through
the detector. Because of edge effects the efficiency, which is, of course, a
very strong function of detector size, is only a few percent after final cuts,
even in the vertical direction. Nevertheless, we will identify background
atmospheric neutrinos and establish meaningful limits on WIMP fluxes from the
center of the Earth.

That this data set, including prefiltering, is relatively well simulated by
the Monte Carlo is shown in Fig.\,15. The results reinforce the conclusions,
first drawn from Figs.~11 and 13, that we understand the performance of the detector. Cuts are on the number of ``direct'' photons, i.e.\ photons which arrive within time residuals of $[-15; 25]$~ns relative to the predicted time. The latter is the time it takes for Cherenkov photons to reach the OM from the reconstructed muon track. The choice of residual reflects the present resolution of our time measurements and allows for delays of slightly scattered photons.
The reconstruction capability of AMANDA-80 is illustrated in Fig.\,16. Comparison of the reconstructed zenith angle distribution of atmospheric muons
and the Monte Carlo is shown in Fig.\,16a for 3 cuts in $N_{direct}$. For
$N_{direct} \geq 5$, the resolution is 2.2~degrees as shown in Fig.\,16b.

\begin{figure}[t]
\centering\leavevmode
\epsfxsize=3.5in\epsffile{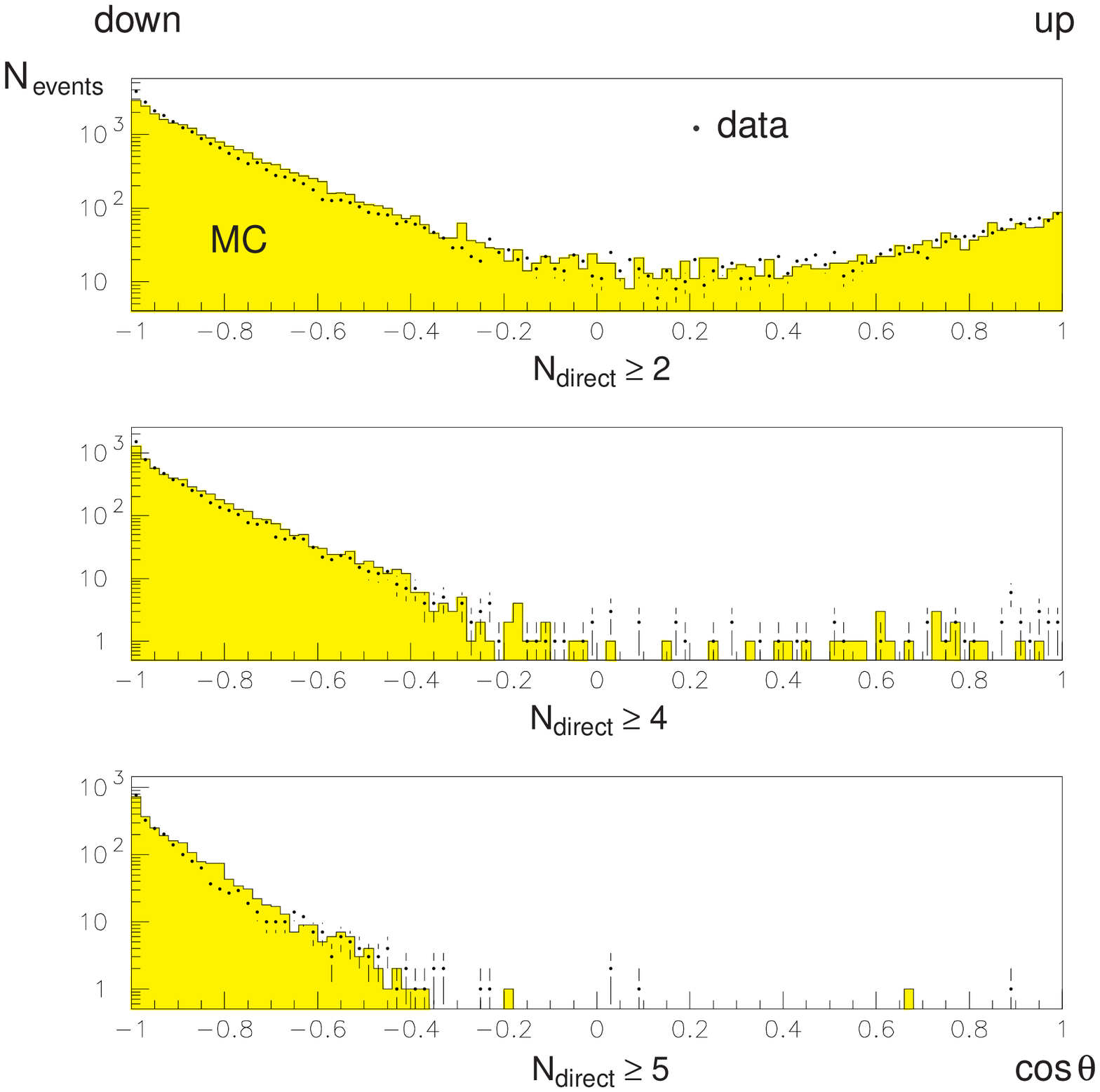}

\caption{$\cos(\theta_{\rm rec})$ is shown with cuts on the number of residuals in the interval $[-15; 25]$ ns. The histogram represents Monte Carlo
simulations with a trigger of 8 or more hits in 2 msec and the dots represent the real data. (Notice that up and down directions are reversed from Fig.\,11)}
\end{figure}

The final cut selecting WIMP candidates requires 6 or more residuals in the
interval $[-15, +15]$~ns and $\alpha \geq 0.1$~m/ns. Here $\alpha$ is the
slope parameter obtained from a plane wave fit $z_i= \alpha t _i+ \beta$, where $z_i$ are the vertical coordinates of hit OMs and $t_i$ the times at which they were hit. The cut selects muons moving vertically along the strings and pointing back towards the center of the Earth.

The two events surviving these cuts are shown in Fig.\,17. Their properties are summarized in Table\,5. The expected number of atmospheric neutrino events
passing the same cuts is $4.8 \pm0.8 \pm1.1$. With only preliminary calibration, the systematic error in the time-calibration of the PMTs is $\sim$15 ns. This reduces the number of expected events to $2.9 \pm0.6 \pm0.6$. The fact that the parameters of both events are not close to the cuts imposed, reinforces their significance as genuine neutrino candidates. Their large tracklengths suggest neutrino
energies in the vicinity of 100~GeV which implies that the parent neutrino
directions should align with the measured muon track to better than 2~degrees. Conservatively, we conclude that we observe 2 events on a background
of 4.8 atmospheric neutrinos. With standard statistical techniques this result can be converted into an upper limit on an excess flux of WIMP origin; see Fig.\,13.

\renewcommand{\thefigure}{\arabic{figure}a}
\begin{figure}[t]
\centering\leavevmode
\epsfxsize=3in\epsffile{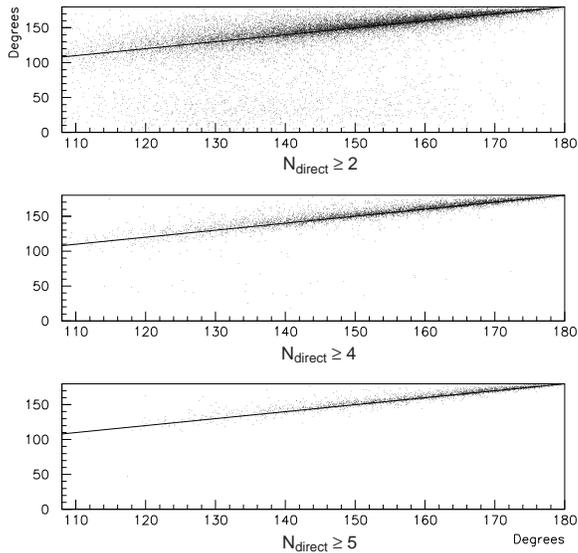}

\caption{Scatter-plot showing the AMANDA-reconstructed $\theta$-angle of
atmospheric muons versus the MC, at several cut levels.}
\end{figure}

\addtocounter{figure}{-1}\renewcommand{\thefigure}{\arabic{figure}b}
\begin{figure}[t]
\centering\leavevmode
\epsfxsize=3in\epsffile{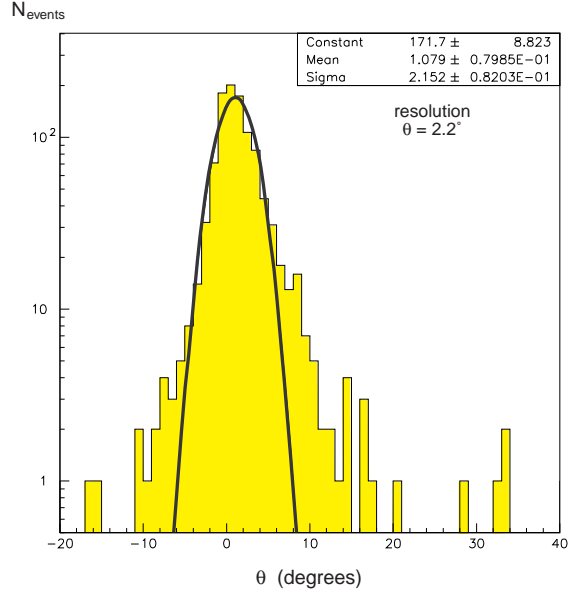}

\caption{$\theta_{\rm rec} - \theta_{\rm MC}$ for reconstructed atmospheric Monte Carlo simulated muons with at least five residuals in the
interval $[-15; 15]$~ns.}
\end{figure}
\renewcommand{\thefigure}{\arabic{figure}}

In order to interpret this result, we have simulated AMANDA-80 sensitivity to
the dominant WIMP annihilation channels\cite{joakim}:
into $b\bar b$ and $W^+ W^-$. The
upper limits on the WIMP flux are shown in Fig.\,18 as a function of the WIMP
mass. Limits below 100~GeV WIMP mass are poor because the neutrino-induced
muons (with typical energy $\simeq m_{\chi}/6$) fall below the AMANDA-80
threshold. For the heavier masses, limits approach the limits set by other
experiments in the vicinity of  $10^{-14}\rm\, cm^{-2}\, s^{-1}$. We have
previously discussed how data, already on tape from AMANDA-300, will make new
incursions into the parameter space of supersymmetric models.

\begin{figure}[t] 
\centering\leavevmode
\epsfysize=4in\epsffile{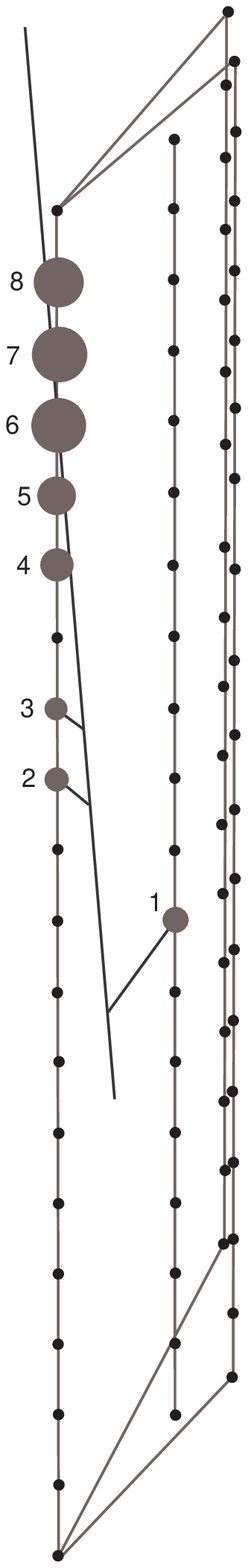}\hspace{1in}
\epsfysize=4in\epsffile{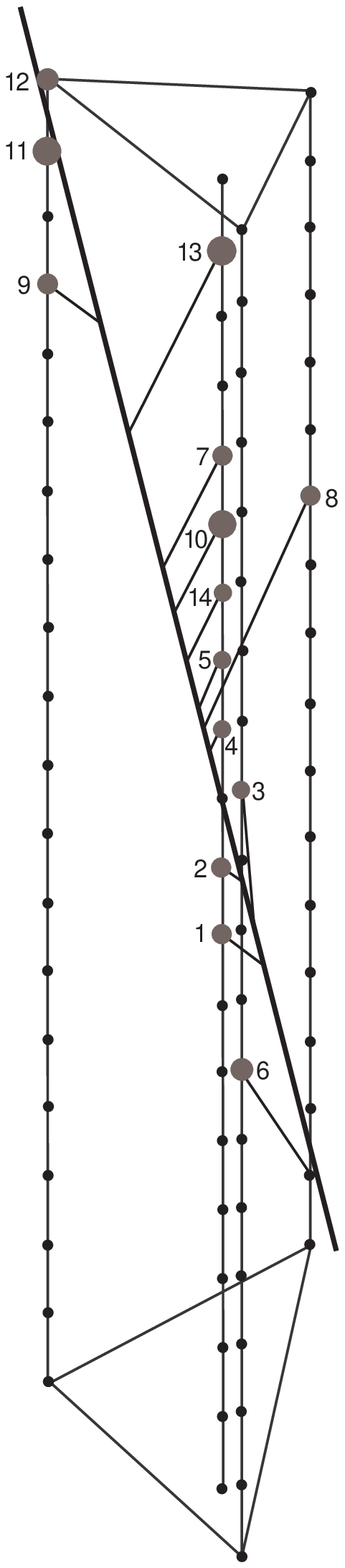}

\caption{Events reconstructed as up-going satisfying the cuts imposed on the data to
search for neutrinos from WIMP annihilation in the center of the Earth.}
\end{figure}

\begin{table}[b]
\begin{center}
\begin{tabular} {|l||c|c|}
\hline
Event ID\# &4706879 & 8427905 \\
\hline
$\alpha$ [m/ns]& 0.19 & 0.37\\
Length [m] &295 & 182\\
Closest approach [m]&2.53  &1.23  \\
$\theta_{rec}$[$^\circ$] &14.1 & 4.6 \\
$\phi_{rec}$[$^\circ$]  &92.0  &  348.7\\
Likelihood/OM &  5.9 &  4.2 \\
OM multiplicity & 14 & 8  \\
String multiplicity & 4  &2  \\
\hline
\end{tabular} \\
\caption{Characteristics of the two events
reconstructed as up-going muons.}
\label{tab:two_events}
\end{center}
\end{table}

\begin{figure}[t]
\centering\leavevmode
\epsfxsize=3.75in\epsffile{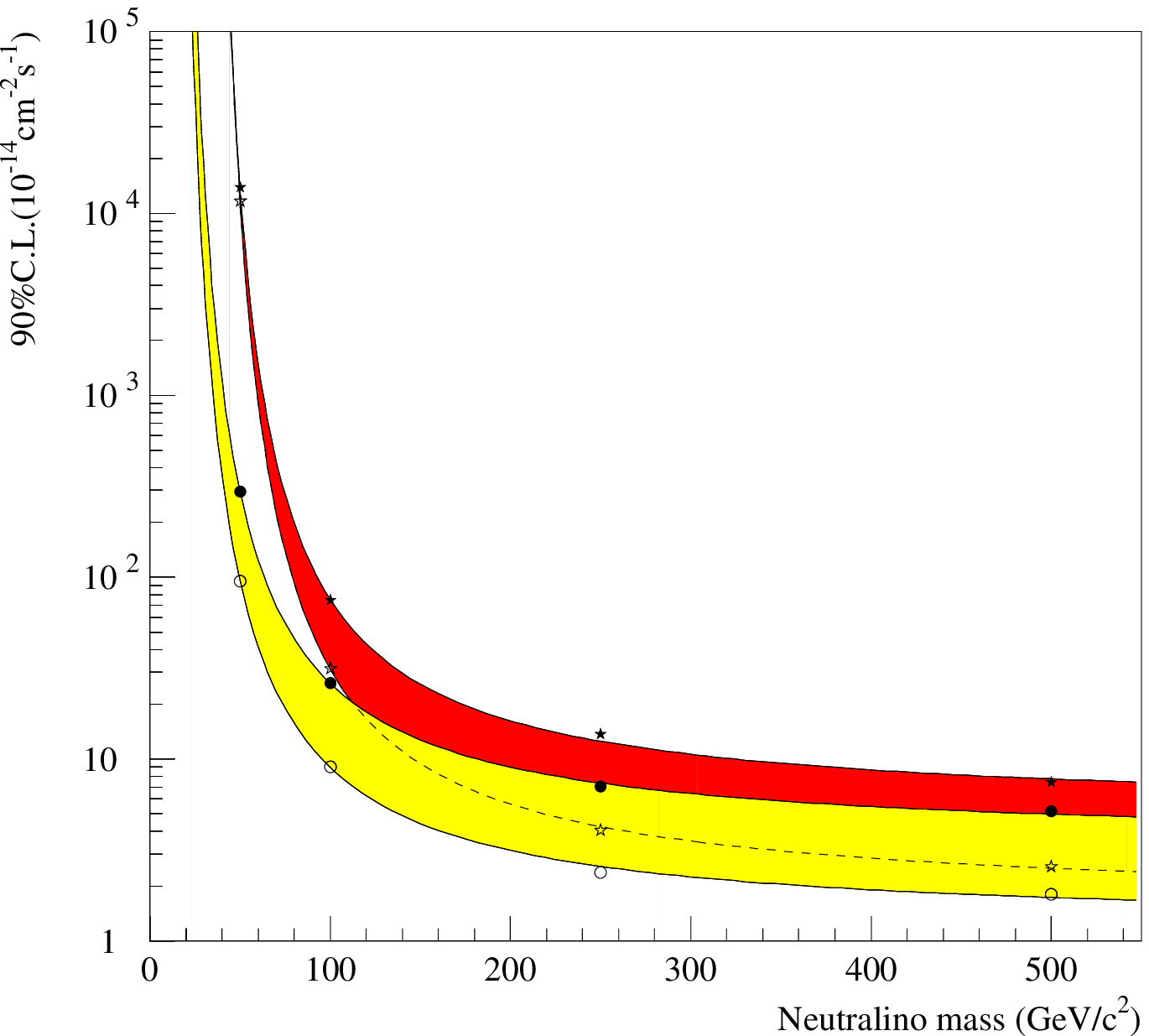}

\caption{Upper limit at the 90\% confidence level on the muon flux from the
center of the Earth as a function of neutralino mass.
The light shaded band represents the $W^+ W^-$ annihilation channel
and the dark one represents the $b\bar b$ annihilation channel. The
width of the bands reflects the inadequate preliminary calculation.}
\end{figure}

\section{Kilometer-Scale Detectors}

\subsection{Towards ICE CUBE(D)}

We concluded in previous sections that the study of AGN and GRB is likely to require the construction of  a kilometer-scale detector. Other arguments support this conclusion\cite{halzen}. A strawman detector with effective area in excess of 1\,km$^2$ consists of 4800\,OM's: 80 strings spaced by $\sim$\,100\,m, each instrumented with 60\,OM's spaced by 15\,m. A cube with a side of 0.8\,km is thus instrumented and a through-going muon can be visualized by doubling the length of the lower track in Fig.\,9a. It is straightforward to convince oneself that a muon of TeV energy and above, which generates single photoelectron signals within 50\,m of the muon track, can be reconstructed by geometry only. The spatial positions of the triggered OM's allow a geometric track reconstruction with a precision in zenith angle of:
\begin{equation}
\mbox{angular resolution} \simeq {\mbox{OM spacing}\over\mbox{length of the track}}
\simeq \mbox{15\,m/800\,m} \simeq \mbox{1\,degree};
\end{equation}
no timing information is required. Timing is still necessary to establish whether a track is up- or down-going, not a challenge given that the transit time of the muon exceeds 2 microseconds.  Using the events shown in Fig.\,9, we have, in fact, already demonstrated that we can reject background cosmic ray muons. Once ICE CUBE(D) has been built, it can be used as a veto for AMANDA and its threshold lowered to GeV energy.

In reality, noise in the optical modules and multiple events, crossing the detector in the relatively long triggering window, will interfere with these somewhat over-optimistic conclusions. This is where an ice detector will be at its greatest advantage however. Because of the absence of radioactive potassium, the background counting rate of OMs deployed in sterile ice can be reduced by close to 2 orders of magnitude. 

With half the number of OM's and half the price tag of the Superkamiokande and SNO solar neutrino detectors, the plan to commission such a detector over 5 years is not unrealistic. The price tag of the default technology used in AMANDA-300 is \$6000 per OM, including cables and DAQ electronics. This signal can be transmitted to the surface by fiber optic cable without loss of information. Given the scientific range and promise of such an instrument, a kilometer-scale neutrino detector must be one of the best motivated scientific endeavors ever.

\subsection{Water and Ice}

The optical requirements of the detector medium can be readily evaluated, at least to first order, by noting that string spacings determine the cost of the detector. The attenuation length is the relevant quantity because it determines how far the light travels, irrespective of whether the photons are lost by scattering or absorption. Remember that, even in the absence of timing, hit geometry yields degree zenith angle resolution. Near the peak efficiency of the OM's the attenuation length is 25--30\,m, larger in deep ice than in water below 4\,km. The advantage of ice is that, unlike for water, its transparency is not degraded for blue Cerenkov light of lower wavelength, a property we hope to take further advantage of by using wavelength-shifter in future deployments.

The AMANDA approach to neutrino astronomy was initially motivated by the low noise of sterile ice and the cost-effective detector technology. These advantages remain, even though we know now that water and ice are competitive as a detector medium. They are, in fact, complementary. Water and ice seem to have similar attenuation length, with the role of scattering and absorption reversed; see Table\,6. As demonstrated with the shallow AMANDA strings\cite{porrata}, scattering can be exploited to range out the light and perform calorimetry of showers produced by electron-neutrinos and showering muons. Long scattering lengths in water may result in superior angular resolution, especially for the smaller, first-generation detectors. This can be exploited to reconstruct events far outside the detector in order to increase its effective volume.

\begin{table}[h]
\def\arraystretch{1.5}\tabcolsep=1em
\caption{Optical properties of South Pole ice at 1750\,m, Lake Baikal water at 1\,km, and the range of results from measurements in ocean water below 4\,km.}
\smallskip
\centering\leavevmode
\begin{tabular}{lccc}
\hline
& (1700 m)&&\\[-1.5ex]
$\lambda = 385$ nm\,$^*$& AMANDA& BAIKAL& OCEAN\\
\hline
attenuation& $\sim 30$ m\,$^{**}$& $\sim 8$ m& 25--30 m\,$^{***}$\\
absorption& $95\pm5$ m& 8 m& ---\\
scattering& $24\pm 2$ m& 150--300 m& ---\\[-2ex]
length&&&\\
\hline
\multicolumn{4}{l}{\llap{$^*$}\,peak PMT efficiency}\\[-1ex]
\multicolumn{4}{l}{\llap{$^{**}$}\,same for bluer wavelengths}\\[-1ex]
\multicolumn{4}{l}{\llap{$^{***}$}\,smaller for bluer wavelengths}
\end{tabular}
\end{table}

\section*{Acknowledgements}

We thank Jaime Alvarez for a careful reading of the manuscript.
This work was supported in part by the University of Wisconsin
Research Committee with funds granted by the Wisconsin Alumni Research
Foundation, and in part by the U.S.\,Department of Energy under Grant
No.\,DE-FG02-95ER40896.

\end{document}